\documentclass[preprint,showpacs,preprintnumbers,amsmath,amssymb]{revtex4}

\usepackage{graphicx}
\usepackage{dcolumn}
\usepackage{bm}

\begin{document}


\title{Magnetic susceptibility of ultra-small superconductor grains}

\author{V. N. Gladilin$^{*}$, V. M. Fomin$^{*,\sharp}$, J. T. Devreese$^{\sharp}$}
\affiliation{ Theoretische Fysica van de Vaste Stoffen (TFVS), Departement
Natuurkunde, Universiteit Antwerpen, Universiteitsplein 1,
B-2610 Antwerpen, Belgium}

\date{\today}

\begin{abstract}

For assemblies of superconductor nanograins, the magnetic response is analyzed as a function of both temperature and magnetic field. In order to describe the interaction energy of electron pairs for a huge number of many-particle states, involved in calculations, we develop a simple approximation, based on the Richardson solution for the reduced BCS Hamiltonian and applicable over a wide range of the grain sizes and interaction strengths at arbitrary distributions of single-electron energy levels in a grain. Our study is focused upon ultra-small grains, where both the mean value of the nearest-neighbor spacing of single-electron energy levels in a grain and variations of this spacing from grain to grain significantly exceed the superconducting gap in bulk samples of the same material. For these ultra-small superconductor grains, the overall profiles of the magnetic susceptibility as a function of magnetic field and temperature are demonstrated to be qualitatively different from those for normal grains. We show that the analyzed signatures of pairing correlations are sufficiently stable with respect to variations of the average value of the grain size and its dispersion over an assembly of nanograins. The presence of these signatures does not depend on a particular choice of statistics, obeyed by single-electron energy levels in grains.

\end{abstract}

\maketitle

\section{Introduction}

Recent intensive theoretical studies of pairing correlations in small superconductor grains (see Ref.~\cite{3} for a review) are inspired, to a great extent, by the experiments of Ralph, Black and Tinkham~\cite{brt1,brt2,brt3}, who measured tunneling spectra of individual nanosize Al grains. For grains with an even number of electrons and the mean value of the nearest-neighbor spacing for single-electron energy levels, $d$, significantly smaller than the bulk superconducting gap $\Delta_b$, the measurements~\cite{brt1,brt2,brt3}  revealed a gap between the ground state and the lowest excited state of a grain. The width of this spectroscopic gap is larger than or comparable with $ 2 \Delta_b$ reflecting the fact that the energy difference between the fully paired ground state and the lowest excited state, which involves one broken pair, is at least as large as the energy cost for pair breaking. The lowest excitation energy contains two contributions: one contribution ($\sim d$) is given by the energy spacing between the corresponding single-electron energy levels and just coincides with the lowest excitation energy for a normal grain, while another ($\sim  2 \Delta_b$) corresponds to the change of the pairing-interaction energy due to pair breaking. When the mean spacing $d$ is much smaller than $\Delta_b$, the lowest excitation energy considerably exceeds $d$ so that a specroscopic gap in the tunneling spectrum can be clearly observed and unambiguously attributed to the presence of pairing correlations in a grain. However, in the case of ultra-small grains with $d>\Delta_b$, the lowest excitation energy is mainly determined by the energy difference between single-electron energy levels and hence pairing correlations can hardly be detected in a convincing way through the tunneling spectra of individual grains. 

For ultra-small grains with $d>\Delta_b$, traces of superconducting correlations in thermodynamic properties (specific heat, magnetic susceptibility) have become a subject of theoretical investigations~\cite{dil2000,cr2000,4,rc2001,fal2002}. 
In Refs.~\cite{dil2000,cr2000,rc2001}, pairing effects in the spin magnetization are studied for a particular case of grains with uniformly spaced single-electron energy levels. Signatures of pairing in the magnetic response of those grains were shown to persist even for $d\gg \Delta_b$. 
In particular, the study~\cite{dil2000} of the magnetic susceptibility in the limit of zero magnetic field revealed that the susceptibility of an odd superconductor grain (for short, we will use the terms `even grain' or `odd grain' for a grain with an even or odd number of electrons, respectively) should have a minimum as a function of temperature $T$ at $k_{\mathrm B}T\sim d$, where $k_{\mathrm B}$ is the Boltzmann constant. This minimum originates from a superposition of the Curie-like contribution of a single electron, which is unpaired at $T=0$, and contributions, which are related to electrons paired at $T=0$ and increase with temperature due to pair breaking. Such a non-monotonous behavior is visible in the calculated magnetic susceptibility~\cite{dil2000} of grains with the ratio $d/ \Delta_b$ as large as 50. 
Calculations of the zero-field magnetic susceptibility have been extended~\cite{fal2002} to assemblies of superconductor grains, assuming that single-electron energy levels in grains follow the Gaussian orthogonal ensemble (GOE) distribution~\cite{rm}. The zero-field magnetic susceptibility of assemblies, which contain only odd grains or approximately equal fractions of even and odd grains, as a function of temperature is shown~\cite{fal2002} to manifest a non-monotonous behavior similar to that described above. The zero-temperature magnetic response of superconductor nanograins with the GOE statistics of single-electron energy levels has been investigated~\cite{4} using a perturbative approach to calculate the interaction energy from the Richardson equations~\cite{RWR1}, which provide an exact solution for the reduced BCS Hamiltonian.    
While for normal grains the zero-temperature magnetic susceptibility monotonously increases with magnetic field $H$, a maximum at $\mu_{\mathrm B}H \sim d$ ($\mu_{\mathrm B}$ is the Bohr magneton) was found in the susceptibility of superconductor grains as a function of $H$. For ultra-small superconductor grains with the dimensionless interaction constant as large as 0.28, the calculated height of this maximum~\cite{4} is by two percent larger than the saturation value of the magnetic susceptibility for both superconductor and normal grains at $H\to \infty$. 

This paper is aimed to study the magnetic response of superconductor grains as a function of both temperature and magnetic field.  
Our study is focused upon ultra-small grains, where both the mean value of the nearest-neighbor spacing of single-electron energy levels in a grain and variations of this spacing from grain to grain exceed the superconducting gap $\Delta_b$ in bulk samples of the same material. For the ultra-small grains, we analyze observable manifestations of pairing correlations in the behavior of the magnetic susceptibility as a function of $T$ and $H$.      
Since magnetization measurements on single nanograins look problematic, we perform calculations for assemblies of grains. 
It is worth noting that an appropriate choice of statistics to describe the energy-level distribution depends on concrete nanostructures.  Thus, ultra-small superconductor grains, prepared by ion implantation, are known to possess a regular geometric shape~\cite{john1,john2,john3}, implying that single-electron energy levels can be degenerate and hence that they do not obey the GOE statistics, often used for superconductor nanograins in the literature.  Below, we consider two cases: (i) parallelepiped-shaped grains~\cite{2}, where single-electron energy levels obey the Poisson statistics and (ii) grains with the GOE statistics of single-electron energy levels. 
In order to make the analysis of the magnetic response possible for large assemblies of nanograins at non-zero temperatures, when a large number of many-particle states are involved in calculations, we develop an approximation for the interaction energy of pairs in a nanograin. This approximation, derived on the basis of the Richardson equations~\cite{RWR1},  combines the advantage of simplicity with the possibility to analyze the effects~\cite{2} related to an extremely high sensitivity of the interaction energy to a specific distribution of single-electron energy levels. 

The rest of this paper is organized as follows.
In Section II we describe the model used to calculate the magnetic response of superconductor nanograins. In Subsection II.A, we consider the Richardson solution for the reduced BCS equation as applied to the energy of pairs in the absence of a magnetic field. Energy levels in nanograins in the presence of a magnetic field and the spin magnetization are analyzed in Subsection II.B. 
In Subsection II.C we develop an approximation for calculating the interaction energy of pairs in a nanograin. In Section III the proposed approximation is used to study the magnetic susceptibility of superconductor nanograins with different parity of the number of electrons as a function of magnetic field and temperature. We examine also the effect of a particular choice of the single-electron energy-level statistics on the magnetic response of superconductor nanograins. Further, we consider a dependence of the magnetic susceptibility on the interaction strength as well as on the grain size and its variations within an assembly of grains. Conclusions are formulated in Section IV.  In Appendix A we analyze -- as an instructive example -- a grain, where in the absence of a magnetic field single-electron states in the interaction band have equal energy. In Appendix B the accuracy of the proposed approximation for the interaction energy is checked by comparing the magnetic susceptibility, calculated on the basis of this approximation, with the numerically exact results for some particular cases.

\section{Formalism}

\subsection{Energy levels at zero magnetic field}

For a nanosize grain, the BCS pairing Hamiltonian can be written as (see, e.~g., Refs.~\onlinecite{jvd96,ml97}) 
\begin{eqnarray}
H = \sum_{i,\sigma}\varepsilon_i a^{\dag}_{i\sigma}a_{i\sigma} -
\lambda d\sum_{i,j\in {\cal I} } a^{\dag}_{i\uparrow}a^{\dag}_{i\downarrow}a_{j\downarrow}
a_{j\uparrow},
\label{HAM}\end{eqnarray}
where the operator $ a^{\dag}_{i\sigma}$ ($a^{ }_{j\sigma}$) creates (annihilates) an electron with the spin $\sigma$ in the $i$-th spin-degenerate single-electron state.  We will assume that the index $i$ ($i=1,2,\ldots$) labels these states in the order of increasing energy $\varepsilon_i$. 
The second term in Eq.~(\ref{HAM}) is the interaction Hamiltonian. The sum in this term is over the set ${\cal I}$ of $I$ states inside the energy interval $(\varepsilon_{\rm F}- \hbar\omega_{\rm D}, \varepsilon_{\rm F}+ \hbar\omega_{\rm D}$), which will be referred to as the interaction band, with $\hbar\omega_{\rm D}$, the Debye energy, and $\varepsilon_{\rm F}$, the Fermi energy. The interaction strength is a product of the mean energy-level spacing within the interaction band, $d=2\hbar\omega_D /I$, and the dimensionless interaction parameter $\lambda$. 

The electrons occupying levels outside the interaction band are straightforwardly described by the first term in the r.h.s. of Eq.~(\ref{HAM}). Therefore, we will focus now on the electrons, which reside in the interaction band. Let the interaction band be populated by $2n+b$ electrons, of which $b$ electrons are on singly occupied levels (the set of these levels, blocked to the pair scattering, is denoted as ${\cal B}$). The electrons on singly occupied levels do not participate in the pair scattering and their contribution to the energy of the $N$-electron system under consideration is simply $\sum_{i\in {\cal B}} \varepsilon_i$. The remaining $2n$ electrons form $n$ pairs, which are distributed over the set ${\cal U}= {\cal I}\setminus {\cal B}$ of $I-b$ unblocked levels. Richardson~\cite{RWR1} showed that the energy of these pairs is given by the sum of $n$ parameters $E_{1}, \ldots, E_{n}$, which are non-degenerate roots ($E_{\mu}\neq E_{\nu}$ for all $\mu\neq\nu$) of the following set of $n$ coupled equations:
\begin{eqnarray}
\frac{1}{\lambda d}-\sum_{j\in {\cal U}}\frac{1}{2\varepsilon_j - E_{\nu}} +
\sum_{\mu=1\atop\mu \neq \nu}^{n}\frac{2}{E_{\mu}- E_{\nu}} =0, \qquad
\nu=1,\ldots ,n\;.
\label{R1}\end{eqnarray}

If the interaction parameter $\lambda$ reduces to zero, each solution of Eq.~(\ref{R1}) smoothly evolves into a certain set of bare pair energies~\cite{RWR1}. In other words, for a given set ${\cal U}$ of unblocked bare energy levels in the interaction band, a specific choice of a subset ${\cal N}$ of $n$ levels, doubly occupied in the limit $\lambda \to 0$, uniquely determines a solution of Eq.~(\ref{R1}) at $\lambda \neq 0 $. Due to such a one-to-one correspondence between many-electron states at $\lambda=0$ and those at $\lambda \neq 0 $, the states in a superconductor grain can be classified according to the limiting distributions of electrons over bare energy levels for $\lambda \to 0$. 

Let the total number of electrons in a grain be $2K+P$, where the non-negative integer $K$ coincides with the number of doubly occupied bare energy levels in the ground state at $\lambda=0$ in the absence of a magnetic field, while the parity parameter $P$ takes the values 0 or 1 for, respectively, an even or odd  grain. For the energy levels of this $(2K+P)$-electron system in the absence of a magnetic field we use the notation $E_{qk}$, 
where $K-k$ gives the number of doubly occupied bare energy levels at $\lambda=0$,  while $q$ labels a specific distribution of $K-k$ doubly occupied levels and $(2k+P)$ singly occupied levels within the set $\{ \varepsilon_i\}$. Let ${\cal D}^{(qk)}$ and ${\cal S}^{(qk)}$ denote, respectively, the sets of doubly and singly occupied energy levels $\varepsilon_i$ for the state with given $q$ and $k$. 
The parameters $q$ and $k$ uniquely determine also the sets ${\cal B}^{(qk)}$, ${\cal U}^{(qk)}$, and ${\cal N}^{(qk)}$ of singly occupied (blocked), unblocked, and doubly occupied (at $\lambda=0$) bare energy levels in the interaction band, as well as the number of levels in these sets: $b^{(qk)}$, $u^{(qk)}=I-b^{(qk)}$, and $n^{(qk)}$, respectively. 
For fixed values of $q$ and $k$, only one solution of equations~(\ref{R1}) reduces to the set of bare energy levels $\{ 2\varepsilon_{i_{\mu}^{(qk)}} \}$ ($\mu=1,\ldots, n^{(qk)}$), where $i_{\mu}^{(qk)}$ are the elements of the set ${\cal N}^{(qk)}$.  The roots, which belong to this solution, can be denoted as $E_{\mu}^{(qk)}$ ($\mu=1,\ldots, n^{(qk)}$). In parallel to the set of  $E_{\mu}^{(qk)}$ ($\mu=1,\ldots, n^{(qk)}$), it is convenient to introduce the equipotent set of parameters $ \tilde E_{i}^{(qk)}$ ($i\in {\cal N}^{(qk)}$), determined by the relation $\tilde E_{i_{\mu}^{(qk)}}^{(qk)}\equiv E^{(qk)}_{\mu}$ ($\mu=1,\ldots, n^{(qk)}$). In the set $\{ \tilde E_{i}^{(qk)}\}$, each root of equations~(\ref{R1}) is labelled in accordance with its limiting value at $\lambda \to 0$, that is $2\varepsilon_i$ ($i\in {\cal N}^{(qk)}$). So, the energy $E_{qk}$ takes the form  
\begin{eqnarray}
E_{qk}= \sum_{i\in {\cal S}^{(qk)}} \varepsilon_i +2\sum_{i\in {\cal D}^{(qk)}} \varepsilon_i - E_{qk}^{\mathrm I} , 
\label{netot}\end{eqnarray}
where the interaction energy for $n^{(qk)}$ electron pairs, which reside in the interaction band, is 
\begin{eqnarray}
E_{qk}^{\mathrm I}=  \sum_{i\in {\cal N}^{(qk)}} \left( 2\varepsilon_i - \tilde E_{i}^{(qk)} \right) 
\label{neint}\end{eqnarray}
with $\tilde E_{i}^{(qk)}$, the roots of the equations
\begin{eqnarray}
\frac{1}{\lambda d}-\sum_{j\in {\cal U}}\frac{1}{2\varepsilon_j - \tilde E_{i}^{(qk)}} +
\sum_{j \in {\cal N}^{(qk)}\atop j \neq i}\frac{2}{\tilde E_{j}^{(qk)}- \tilde E_{i}^{(qk)}} =0 \,, \qquad
i \in {\cal N}^{(qk)}\;.
\label{R1n}\end{eqnarray}

Let us label with $q=0$ the lowest $(2K+P)$-electron state with a given number $k$. In this state at $\lambda \to 0$, $2(K-k)$ electrons fully occupy the lowest $(K-k)$ energy levels $\varepsilon_i$, while the next $(2k+P)$ energy levels $\varepsilon_i$ are singly occupied by the remaining $(2k+P)$ electrons. The corresponding energy is         
\begin{eqnarray} 
E_{0k}=2\sum\limits_{i=1}^{K-k} \varepsilon_i+\sum\limits_{i=K-k+1}^{K+k+P} \varepsilon_i -E_{0k}^{\mathrm{I}}.  
\label{en0}
\end{eqnarray} 
It is obvious that the interaction energy $E_{0k}^{\mathrm{I}}$ differs from zero only for $k<n^{(00)}$, where $n^{(00)}$ is the number of electron pairs in the ground state, which has the energy  
\begin{eqnarray} 
E_{00}=2\sum\limits_{i=1}^K \varepsilon_i+P\varepsilon_{K+1}-E_{00}^{\mathrm{I}}.
\label{eng}
\end{eqnarray} 
From Eq.~(\ref{en0}), one can easily derive the relation
\begin{eqnarray} 
{\Delta_k  E_{0k}}= -\Delta_k  E_{0k}^{\mathrm{I}}+\left(\varepsilon_{K+k+P+1}-\varepsilon_{K-k}\right),  
\label{en1}
\end{eqnarray} 
where the forward difference of any function $F_k$ of an integer variable $k$ is defined as $\Delta_k F_{k} \equiv F_{k+1}-F_{k}$. 
It is worth mentioning that for $k<n^{(00)}$ the quantity $(-\Delta_k  E_{0k}^{\mathrm{I}})$, which has the meaning of the pair-breaking energy, is strictly positive. Therefore, as follows from Eq.~(\ref{en1}), the energy $E_{0k}$ is a strictly increasing function of $k$ for $k<n^{(00)}$. (At $k\geq n^{(00)}$ the forward difference $\Delta_k  E_{0k}^{\mathrm{I}}$ may take zero value if the energy levels $\varepsilon_i$ are degenerate.)

\subsection{Spin magnetization}

When analyzing the magnetic response of nanograins, we assume these grains to be flat and subjected to a magnetic field parallel to the grain base. For an in-plane magnetic field, the orbital magnetization of electrons in a flat nanograin can be neglected \cite{bdr97,salin,bd99}. In a magnetic field $H$, the energy of a $(2K+P)$-electron state depends on the spin orientation of electrons on singly occupied energy levels $\varepsilon_i$:  
\begin{eqnarray} 
E_{qkl}=E_{qk} -\left(2k+P-2l\right)\mu_{\mathrm B}H.  
\label{en}
\end{eqnarray} 
Here $l$ ($l=0,...,2k+P$) is the number of spin-down electrons on singly occupied energy levels $\varepsilon_i$ (we use the terms `spin-up' and `spin-down' for electrons with spins, which are, respectively, parallel and antiparallel to the applied magnetic field), and the Land\'e factor for the electron spin is assumed to equal 2. The spin magnetization of electrons in a state with the energy~(\ref{en}), does not depend on $q$:  
\begin{eqnarray} 
M_{kl}=\left(2k+P-2l\right)\mu_{\mathrm B}.  
\label{mag0}
\end{eqnarray} 

At a temperature $T$, the spin magnetization of the grain can be expressed as      
\begin{eqnarray} 
M=\frac{1}{Z} \sum\limits_{k=0}^K \sum\limits_{l=0}^{2k+P} C^{2k+P}_{l} M_{kl} \sum\limits_q\exp \left( -\frac{
E_{qkl} }{k_{\mathrm B}T}\right),  
\label{mag}
\end{eqnarray} 
with the partition function
\begin{eqnarray} 
Z=\sum\limits_{k=0}^K \sum\limits_{l=0}^{2k+P}C^{2k+P}_{l} \sum\limits_q \exp \left( -\frac{
E_{qkl} }{k_{\mathrm B}T}\right).  
\label{Z}
\end{eqnarray} 
The presence of the binomial coefficients $C^{2k+P}_{l}$ in Eqs.~(\ref{mag}) and (\ref{Z}) takes into account the degeneracy of the energy level $E_{qkl}$ with respect to permutations between spin-up and spin-down electrons on singly occupied energy levels $\varepsilon_i$. 

To facilitate the further analysis, let us first consider magnetization of a nanograin at zero temperature. At $T=0$, magnetization is determined by the polarization of electron spins in the ground state. To find the ground state at a given magnetic field, one has to minimize the energy $E_{qkl}$ with respect to the numbers $q$, $k$, and $l$. As obvious from Eq.~(\ref{en}), in the ground state the number of spin-down electrons on singly occupied levels $\varepsilon_i$ must be zero: $l=0$. We also note that only the first term in the r.h.s. of Eq.~(\ref{en}) does depend on $q$. At a fixed number $k$, the minimal value of this term is $E_{0k}$ [see Eq.~(\ref{en0})]. Therefore, the magnetization of the grain at $T=0$ is given by the value $M_{k0}$ with the number $k$, which minimizes the energy     
\begin{eqnarray} 
E_{0k0}=E_{0k} -\left(2k+P\right)\mu_{\mathrm B} H\,.  
\label{enm}
\end{eqnarray} 
Since the energy $E_{0k}$ has been shown to be a non-decreasing function of $k$ (a strictly increasing function of $k$ for $k<n^{(00)}$), at $H\to 0$ the number of singly occupied energy levels $\varepsilon_i$ in the ground state goes to $P$ and, correspondingly, the magnetization is given by the expression $M_{00}=\mu_{\mathrm B}P$. However, as seen from Eq.~(\ref{enm}) with rising $H$ an increase of $k$ can become energetically favorable. Each jump of the value $k_m$, which minimizes the energy $E_{0k0}\left( H\right)$, results in an abrupt increase of the magnetization. 

Under the condition $\Delta_k^2   E_{0k0} \geq 0$ or, equivalently,
\begin{eqnarray} 
{\Delta_k^2 E_{0k}} \geq 0, 
\label{crit}
\end{eqnarray} 
the energy $E_{0k0}\left( H\right)$ as a function of $k$ has only one minimum on the interval $[0,K]$, while the value $k_m$ is the minimum integer $k$, which obeys the inequality  ${\Delta_k E_{0k0}}> 0$. 
In this case, an increase of $H$ results in consecutive jumps of the value $k_m$ from $k$ to $k+1$, where $k=0,1,2,\ldots$.  The magnetic field $H_k$, which corresponds to the change of $k_m$ from $k$ to $k+1$ is determined by the equation ${\Delta_k E_{0k0}}=0$. From Eq.~(\ref{enm}), we find 
\begin{eqnarray} 
{\Delta_k E_{0k0}}={\Delta_k E_{0k}}-2\mu_{\mathrm B}H   
\label{dehk}
\end{eqnarray} 
and, using Eq.~(\ref{en1}),
\begin{eqnarray} 
H_k&&=
\frac{1}{2\mu_{\mathrm B}} \left( -{\Delta_k E_{0k}^{\mathrm{I}}}+\varepsilon_{K+k+P+1}-\varepsilon_{K-k} \right).   
\label{field}
\end{eqnarray} 
The magnetic field $H_k$ given by Eq.~(\ref{field}) is a non-decreasing function of $k$, provided that the inequality~(\ref{crit}) is obeyed. From Eq.~(\ref{en1}) we obtain the equation    
\begin{eqnarray} 
{\Delta_k^2 E_{0k}} = -{\Delta_k^2 E_{0k}^{\mathrm{I}}}+\left( \varepsilon_{K-k}-\varepsilon_{K-k-1}\right)+\left( \varepsilon_{K+k+P+2}-\varepsilon_{K+k+P+1}\right), 
\label{d2Edk2}
\end{eqnarray} 
which clearly shows that inequality~(\ref{crit}) is always satisfied for normal grains, where the interaction energy is identically zero. However, for superconductor grains, the sign of ${\Delta_k^2 E_{0k}}$ is not obvious. On the one hand, for $k<n^{(00)}$, an increase of $k$ reduces the number of electron pairs $n^{(0k)}$ and, simultaneously, enlarges the number $b^{(0k)}$ of singly occupied energy levels $\varepsilon_i$ in the interaction band, which are blocked to pair scattering. 
Both a decrease of $n^{(0k)}$ and an increase of $b^{(0k)}$ weakens the pairing interaction. For this reason, the pair-breaking energy $(-{\Delta_k E_{0k}^{\mathrm{I}}})$ decreases with increasing $k$ and, correspondingly, the values of $(-{\Delta_k^2 E_{0k}^{\mathrm{I}}})$ are negative. On the other hand, for sufficiently small values of the energy differences $\left( \varepsilon_{K-k}-\varepsilon_{K-k-1}\right)$ and $\left( \varepsilon_{K+k+P+2}-\varepsilon_{K+k+P+1}\right)$, the sign of ${\Delta_k^2 E_{0k}}$ in Eq.~(\ref{d2Edk2}) is just the same as the sign of $(-{\Delta_k^2 E_{0k}^{\mathrm{I}}})$. Thus, depending on a specific energy spectrum $\{\varepsilon_i\}$ as well as on particular values of $\lambda$ and $k$, the inequality~(\ref{crit}) may be violated in superconductor grains (an instructive example, which corresponds to the case of $\Delta_k^2 E_{0k}< 0$, is given in Appendix A). A violation of the condition~(\ref{crit}) means that the energy $E_{0k0}$ as a function of $k$ can have two or more minima on the interval $[0,K]$. Since the relative depth of those minima depends on the magnetic field, with rising $H$ an abrupt increase of $k_m$ by a value $\Delta k>1$ is possible. 
As clear from Eq.~(\ref{dehk}) and the definition of $H_k$, a given value $k$ can correspond to the ground state only for $H \leq H_k$. 
Therefore, an abrupt change of $k_m$ from $k_m=k$ to $k_m=k+\Delta k$ may happen only at a magnetic field $H_{k,\Delta k}$, which satisfies the inequality 
\begin{eqnarray} 
H_{k,\Delta k}\leq H_k.
\label{hh}
\end{eqnarray}

As follows from the above considerations, at $T=0$ the magnetization $M$ of a single nanograin forms a staircase, each jump of $M$ resulting in a $\delta$-like peak of the magnetic susceptibility $\chi$. From Eqs.~(\ref{field}) and (\ref{hh}), it is obvious that positions of these peaks are strongly affected by a specific shape-dependent configuration of the energy levels in the grain.  Since this configuration is hardly controlled in experiment, a possibility to convincingly reveal pairing correlations through the susceptibility measurements for a {\it single nanograin} looks unlikely (apart from obvious difficulties of those measurements as such). Therefore, we analyze below the magnetic susceptibility for {\it assemblies of grains}, where some general trends related to the pairing interaction can be manifested. 

For an assembly of grains, we define the magnetic susceptibility as the forward difference quotient
\begin{eqnarray} 
\langle \chi^* \rangle=\frac{ \langle M(H+\Delta H)\rangle -\langle M(H)\rangle }{\Delta H}, 
\label{sus}
\end{eqnarray} 
where angular brackets $\langle \ldots \rangle$ stand for the average over the assembly of grains. The above definition of the magnetic susceptibility is more convenient than the derivative $\langle \chi \rangle=\partial \langle M \rangle/\partial H$ because at $T \to 0$ the latter becomes a set of $\delta$-like peaks for any {\it finite} number of grains. In the calculations below, the values $\Delta H \sim 10^{-2}\langle d \rangle/\mu_{\mathrm B}$ are used. We introduce also the normalized magnetic susceptibility 
\begin{eqnarray} 
\bar\chi \equiv \frac{\langle \chi^* \rangle\langle d \rangle}{ 2\mu_{\mathrm B}^2 }\,. 
\label{susnorm}
\end{eqnarray}

\subsection{Approximation for the interaction energy}

An exact solution of Eqs.~(\ref{R1n}) can be obtained only by numerical computation, which becomes more and more time-consuming with increasing the number of single-particle energy levels and the number of electrons in the interaction band. At the same time, calculations of the spin magnetization for a nanograin require to compute interaction energies for many-electron states, whose number rapidly increases with increasing temperature. Under those circumstances, an exact calculation of the spin magnetization seems problematic, especially for large assemblies of grains. Because of that, here we introduce an approximation, which combines the advantage of simplicity with the possibility to catch the effects~\cite{2} related to the extremely high sensitivity of the interaction energy to the distribution of single-electron energy levels over the interaction band. 

In this subsection, we focus our analysis on the system of the electron pairs distributed over a fixed set ${\cal U}$ of $u$ unblocked spin-degenerate energy levels $\varepsilon_i$ in the interaction band. So, when discussing, e.~g., the ground state (excited states) of the system, we mean the lowest state of these electron pairs (their collective excitations) irrespective of the state of other electrons, contained in the grain. For brevity, we do not explicitly indicate the indices $q$ and $k$, which label many-electron states of the whole grain. Unless otherwise stated, we assume to deal with any of those states.     

We will construct our approximation for the interaction energy as an ``interpolation'' between the results for the limiting cases of weak and strong pairing interaction. 

\subsubsection{Weak interaction}

The case of weak interaction is defined by the inequality 
\begin{eqnarray}
\lambda d \ll {\left|\varepsilon_i - \varepsilon_j \right|}, \quad {\rm for}\quad i,j \in {\cal U},\ i\neq j. 
\label{weak}\end{eqnarray}
According to Eq.~(\ref{R1n}), we have  
\begin{eqnarray}
2\varepsilon_i - \tilde E_{i}=\lambda d\left[1-\sum_{j\in {\cal U} \atop j \neq i}\frac{\lambda d}{2\varepsilon_j - \tilde E_{i}} -
\sum_{j\in {\cal N}\atop j \neq i}\frac{2\lambda d}{\tilde E_{j}- \tilde E_{i}}\right]^{-1} , \qquad
i \in {\cal N}\;.
\label{R11}\end{eqnarray}
Under condition (\ref{weak}), all the roots $\tilde E_{j}$ ($j\in {\cal N}$) in the r.~h.~s. of Eq.~(\ref{R11}) can be approximated~\cite{4} by their limiting values at $\lambda \to 0$. Then the interaction energy (\ref{neint}) becomes 
\begin{eqnarray}
E^{\mathrm I}=  \sum_{i\in {\cal N}}\lambda_i d, 
\label{eintweak}\
\end{eqnarray}
where
\begin{eqnarray}
\lambda_i={\lambda}\left[1-{\lambda d}\left(\sum_{j\in {\cal U}\atop j \neq i}\frac{1}{2(\varepsilon_j - \varepsilon_i)} -
\sum_{j\in {\cal N}\atop j \neq i}\frac{1}{\varepsilon_j- \varepsilon_i}\right)\right]^{-1}.   
\label{alpha}
\end{eqnarray}
Formally, equation (\ref{eintweak}) looks as a sum of interaction energies for $n$ independent subsystems, where each subsystem consists of a single spin-degenerate energy level with a single electron pair on it, the pairing interaction for the $i$-th subsystem being characterized by the interaction parameter $\lambda_i$. The dependence of the interaction energy on a specific choice of a many-electron state is straightforwardly taken into account through the dependence of $\lambda_i$ on the sets ${\cal U}$ and ${\cal N}$ specific for each many-electron state.  

\subsubsection{Strong interaction}

The case of strong interaction is defined by the inequality
\begin{eqnarray}
\lambda d\gg \delta
\label{strong}\end{eqnarray}
with $\delta= 2\max_{i,j\in {\cal U}}|\varepsilon_i-\varepsilon_j|$. Inequality~({\ref{strong}) can be satisfied even at $\lambda<1$ provided that the energy levels $\varepsilon_j$ in the interaction band are (quasi-) degenerate. Under the condition~({\ref{strong}), the roots $\tilde E_{i}$ ($i \in {\cal N}$) as a function of $\lambda d$ demonstrate two qualitatively different types of behavior. While for the ground state all these roots are characterized by $ |2\varepsilon_i-\tilde E_{i}|\sim \lambda d $ and tend to infinity at $\lambda d\to \infty$, for excited states a subset ${\cal G}\subset {\cal N}$ of $g$ roots remain finite at $\lambda d\to \infty$: $|2\varepsilon_j-\tilde E_{i}|< \delta $ for $i\in {\cal G}$ and $j\in {\cal U}$.  The roots, which belong to the subset ${\cal G}$, are shown to represent bosonic pair-hole excitations of the system~\cite{rom}. The algorithm, developed in Ref.~\onlinecite{rom}, allows one to count the number of those roots depending on a specific choice of a subset ${\cal N}$ within the set ${\cal U}$ of unblocked energy levels in the interaction band. 

Taking into account the inequalities $2|\bar\varepsilon_{\cal U}-\varepsilon_j|\leq \delta$ ($j\in {\cal U}$) and $|2\bar\varepsilon_{\cal U}-\tilde E_{i}|\leq \delta$ ($i\in {\cal G}$) with $\bar \varepsilon_{\cal U} = u^{-1}\sum_{j \in {\cal U}}\varepsilon_j$, it is convenient to rewrite  
Eqs.~(\ref{R1n}) for $i \in {\cal N}\setminus {\cal G}$ in the form
\begin{eqnarray}
\frac{1}{\lambda d}-\frac{u-2g}{2\bar \varepsilon_{\cal U} - \tilde E_{i}}
+\sum_{j \in {\cal N}\atop j \neq i}\frac{2}{\tilde E_{j}- \tilde E_{i}}
+\frac{2m_1}{(2\bar \varepsilon_{\cal U} - \tilde E_{i})^2}
+\sum_{p=2}^{\infty}\frac{2m_p-\mu_p}{(2\bar \varepsilon_{\cal U} - \tilde E_{i})^p}=0, \quad i\in {\cal N}\setminus {\cal G},
\label{R1nexp}\end{eqnarray}
where
\begin{eqnarray}
\mu_p=\sum_{j \in {\cal U}}\left(2\bar \varepsilon_{\cal U}- 2\varepsilon_j\right)^p,
\label{mup}\end{eqnarray}
\begin{eqnarray}
m_p=\sum_{j \in {\cal G}}\left(2\bar \varepsilon_{\cal U}- E_j\right)^p.
\label{mp}\end{eqnarray}
Using Eq.~(\ref{R1nexp}), we obtain for the interaction energy the expression
\begin{eqnarray}
E^{\mathrm I}=\lambda d\left[(n-g)(u-n-g+1)-\frac{2n(\bar \varepsilon_{\cal U}-\bar \varepsilon_{\cal N})}{\lambda d}+\frac{(1-2s_1)m_1}{\lambda d}
+\sum_{p=2}^{\infty}\frac{s_p(\mu_p-2m_p)}{(\lambda d)^p}\right]
\label{Eis0}\end{eqnarray}
with $\bar \varepsilon_{\cal N} = n^{-1}\sum_{j \in {\cal N}}\varepsilon_j$ and 
\begin{eqnarray}
s_p=\sum_{j \in {\cal N}\setminus {\cal G}}\left(\frac{\lambda d}{2\bar \varepsilon_{\cal U} - \tilde E_{i}}\right)^p.
\label{spn}\end{eqnarray}
A set of equations for calculating the coefficients $s_p$ is derived by   
summing up Eqs.~(\ref{R1nexp}), multiplied by $(2\bar \varepsilon_{\cal U} - \tilde E_{i})^{-r}$ ($r=0,1,2,\ldots$), over $i\in {\cal N}$. Thus, omitting terms $\sim\delta^2/(\lambda d)^2$ and $\sim\delta/(\lambda d)$ in the sums with $r=0$ and $r=1$, respectively, we obtain
\begin{eqnarray}
n-g-(u-2g)s_1+\frac{2s_2m_1}{\lambda d}=0,
\label{s1n}\end{eqnarray}
\begin{eqnarray}
s_1-s_1^2-(u-2g-1)s_2=0.
\label{s2n}\end{eqnarray}
With $s_1$ and $s_2$ determined by Eqs.~(\ref{s1n}) and (\ref{s2n}), the interaction energy becomes
\begin{eqnarray}
E^{\mathrm I}=&&\lambda d\left[(n-g)(u-n-g+1)-\frac{2n(\bar \varepsilon_{\cal U}-\bar \varepsilon_{\cal N})}{\lambda d}+\frac{(u-2n)m_1}{(u-2g)\lambda d}
\right. \nonumber\\&&
\left.+\frac{(n-g)(u-g-n)(\mu_2-2m_2-2m_1^2)}{(u-2g)^2(u-2g-1)(\lambda d)^2}+\ldots\right],
\label{Eis1}\end{eqnarray}
where the omitted terms are of the order of $\sim\delta^3/(\lambda d)^3$ as compared to the first term in square brackets.
Equation~(\ref{Eis1}) extends the results for the energy of $n$ electron pairs on a single $u$-fold degenerate energy level in the interaction band (see, e.~g., Refs.~\onlinecite{bgk00,bg01}) to the case when this level is quasi-degenerate. 
It is noteworthy that in the important particular case of half filling ($u=2n$) a dependence of the interaction energy on the values $\tilde E_{i}$ with $i\in {\cal G}$ (which in general cannot be found analytically) appears only in the terms of the order of $\sim\delta^p/(\lambda d)^p$ ($p\geq 2$) as compared to the leading term.

\subsubsection{Arbitrary interaction}

When deriving an approximation applicable for arbitrary values of the ratios $\lambda d/(\varepsilon_i-\varepsilon_j)$ ($i,j\in {\cal U}$), we start with the ground state of $n$ pairs, when the set ${\cal N}$ corresponds to the lowest energy levels from the set ${\cal U}$.
For each $i\in {\cal N}$ we define a subset ${\cal U}_i\subset {\cal U}$ of $u_i$ energy levels $\varepsilon_j$ in accordance with the conditions
\begin{eqnarray}
\left|\varepsilon_i - \varepsilon_j \right|\leq c  \quad {\rm for}\quad j \in {\cal U}_i,  
\label{ui}\end{eqnarray}
\begin{eqnarray}
\left|\varepsilon_i - \varepsilon_j \right| > c  \quad {\rm for}\quad j \notin {\cal U}_i.
\label{notui}\end{eqnarray}
The way we choose the value of the ``discrimination parameter'' $c$ will be specified later [see the text after Eq.~(\ref{tillam})]. 
Let ${\cal N}_i$ denote the subset of $n_i$ doubly occupied levels within the set ${\cal U}_i$: ${\cal N}_i={\cal N} \cap {\cal U}_i$. Then equations~(\ref{R1n}) can be rewritten as 
\begin{eqnarray}
\frac{1}{\lambda d}&&-\frac{1}{2\varepsilon_i - \tilde E_{i}}\left(u_i-
\sum_{j \in {\cal U}_i\atop j \neq i}
\frac{2\varepsilon_j-2\varepsilon_i}{2\varepsilon_j - \tilde E_{i}}\right)
+\frac{2}{2\varepsilon_i - \tilde E_{i}}\left(n_i-1+  
\sum_{j \in {\cal N}_i\atop j \neq i} \frac{4\varepsilon_i-\tilde E_{i}-\tilde E_{j}}{\tilde E_{j} - \tilde E_{i}}\right)
\nonumber\\
&&-\sum_{j \in {\cal U}\setminus {\cal U}_i}\frac{1}{2\varepsilon_j - \tilde E_{i}}+\sum_{j \in {\cal N}\setminus{\cal N}_i}\frac{2}{\tilde E_{j} - \tilde E_{i}}=0,
\qquad i\in {\cal N}.
\label{R1appr}\end{eqnarray}
When raising Eq.~(\ref{notui}) to an argument for applicability of a perturbative approach similar to that considered in Subsection II.C.1, the differences $2\varepsilon_j-\tilde E_{i}$ and $\tilde E_{j}-\tilde E_{i}$ ($j\notin {\cal U}_i$), which enter the last two terms in the l.~h.~s. of Eq.~(\ref{R1appr}), can be approximated by their limiting values at $\lambda \to 0$. 
Summing up Eqs.~(\ref{R1appr}), multiplied by $2\varepsilon_i-\tilde E_{i}$, over $i\in {\cal N}$, we find for the ground-state interaction energy: 
\begin{eqnarray}
E^{\mathrm I}_{\rm g.s.}=&&\sum_{i \in {\cal N}}\tilde\lambda_id\left(u_i-n_i+1\right)+\sum_{i \in {\cal N}}\tilde\lambda_i d
\sum_{j \in {\cal U}_i\atop j \neq i}\frac{2\varepsilon_j-2\varepsilon_i}{2\varepsilon_j - \tilde E_{i}}
\nonumber\\
&&+\frac{d}{2}\sum_{i,j \in {\cal N}_i \atop j \neq i}
\frac{(\tilde E_{j}+\tilde E_{i})(\tilde\lambda_i-\tilde\lambda_j)
-4(\tilde\lambda_i\varepsilon_i-\tilde\lambda_j\varepsilon_j)}{\tilde E_{j}-\tilde E_{i}}, 
\label{Eiapp}\end{eqnarray}
where
\begin{eqnarray}
\tilde\lambda_i={\lambda}\left[1-{\lambda d}\left(\sum_{j\in {\cal U}\setminus{\cal U}_i }\frac{1}{2(\varepsilon_j - \varepsilon_i)} -
\sum_{j\in {\cal N}\setminus{\cal N}_i}\frac{1}{\varepsilon_j- \varepsilon_i}\right)\right]^{-1}.   
\label{tillam}
\end{eqnarray}
When treating Eq.~(\ref{ui}) as a criterion of (quasi-) degeneracy of single-electron states, which belong to the set ${\cal U}_i$, the inequalities $2|\varepsilon_j-\varepsilon_i|\ll |2\varepsilon_j-\tilde E_{i}|$ and $2|\varepsilon_j-\varepsilon_i|\ll |\tilde E_{j}-\tilde E_{i}|$ are satisfied for $i,j\in {\cal U}_i$ [cp. the analysis of the ground-state interaction energy under condition~(\ref{strong}) in Subsection II.C.2]. Therefore, the last two terms in the r.~h.~s. of Eq.~(\ref{Eiapp}) can be neglected as compared to the first term. Thus, we obtain the expression   
\begin{eqnarray}
E^{\mathrm I}_{\rm g.s.}= &&\sum_{i \in {\cal N}}\frac{\tilde\lambda_idn_i\left(u_i-n_i+1\right)}{n_i}. 
\label{Eiapp2}\end{eqnarray}

Expression~(\ref{Eiapp2}) corresponds to the approximation of (non-degenerate) energy levels $\varepsilon_j$ ($j\in {\cal U}_i$) by a single degenerate level. As implied from Eq.~(\ref{Eis1}) taken at $g=0$, such an approximation  overestimates the ground-state interaction energy. Obviously, the extent of this overestimation increases with increasing $c$. On the other hand, an underestimate of the parameter $c$ results in oversized values of $\tilde\lambda_i$, determined by the perturbative expression~(\ref{tillam}). As seen from Eq.~(\ref{tillam}), these values have a tendency to diverge for $c\sim\lambda d$. For the above reasons, equation~(\ref{Eiapp2}) is believed to provide an upper bound for the ground-state interaction energy. Therefore, it is natural to choose the parameter $c$ by minimizing the value $E^{\mathrm I}_{\rm g.s.}$ given by Eq.~(\ref{Eiapp2}).        

The ground-state interaction energy, described by Eq.~(\ref{Eiapp2}), can be considered as a sum of the interaction energies for $n$ mutually independent systems weighted with the factors $n_i^{-1}$. Each of these systems contains $n_i$ electron pairs, distributed over a set ${\cal U}_i$ of degenerate single-electron states, and is characterized by the interaction parameter $\tilde\lambda_i$. Assuming that the above structure of the expression for the interaction energy keeps for all the $n$-pair states, we extend our approach to the case of collective excitations of electron pairs by approximating the interaction energy with the expression 
\begin{eqnarray}
E^{\mathrm I}=&&\sum_{i \in {\cal N}}\frac{\tilde\lambda_id(n_i-g_i)(u_i-n_i-g_i+1)}{n_i}. 
\label{Eiafin}\end{eqnarray}
In Eq.~(\ref{Eiafin}), the number of elementary collective excitations $g_i$ is calculated in accordance with the algorithm~\cite{rom} applied separately to each set ${\cal U}_i$ with the subset ${\cal N}_i$ of doubly occupied levels. 
In the limit $\lambda \to \infty$, when the inequality~(\ref{strong}) is satisfied, the sets ${\cal U}_i$ (${\cal N}_i$) for all $i\in {\cal N}$ coincide with ${\cal U}$ (${\cal N}$) so that the r.~h.~s. of Eq.~(\ref{Eiafin}) reproduces  the leading term of the result given by Eq.~(\ref{Eis1}). In the opposite case described by Eq.~(\ref{weak}), both the sets ${\cal U}_i$ and ${\cal N}_i$ reduce to a single element, $i$, and hence equation~(\ref{Eiafin}) transforms into the perturbative result~(\ref{eintweak}).  

In order to assess the accuracy of the proposed approximation, we compare in Fig.~\ref{fig1} the exact and approximate results for the condensation energy,
\begin{eqnarray}
E^{\mathrm C}_{\rm g.s.}=E^{\mathrm I}_{\rm g.s.}-n\lambda d, 
\label{Econ}\end{eqnarray}
in the ground state of a nanograin. The last term in Eq.~(\ref{Econ}) describes the interaction energy for the uncorrelated Fermi ground state. The calculations are performed for parallelepiped-shaped nanograins with a fixed even number of electrons, $2K=4000$, and various shape, described by the aspect ratio $ l_x: l_y: l_z =1: (1+\eta): (1+2\eta)$. The absolute values of $l_i$ ($i=x,y,z$) are determined by  the equation $2K=\langle l_x \rangle \langle l_y \rangle \langle l_z \rangle n_e$, where $n_e$ is the electron density. The single-electron energy spectrum in a hard-wall parallelepiped with sizes $l_x$, $l_y$ and $l_z$ is given by the expression   
\begin{eqnarray}
\varepsilon_i=\frac{\pi^2\hbar^2}{2m}\left(\frac{ n_x^2(i)}{l_x^2} +\frac{ n_y^2(i)}{l_y^2} +\frac{ n_z^2(i)}{l_z^2} \right), \label{eps}
\qquad n_x(i),n_y(i),n_z(i)=1,2,3,...,
\end{eqnarray}
where $m$ is the effective mass of an electron. The correspondence between $i$ and  $n_x(i)$, $n_y(i)$, $n_z(i)$ is determined by the requirement $\varepsilon_i \leq\varepsilon_{i+1}$, which we impose. 
Henceforth, the values~\cite{poole} of $m$ and $n_e$ for Al are used: $m=1.4m_e $ and $n_e=181$~nm$^{-3}$. 
In the ultra-small grains under consideration, the spacing $d$ varies from 1.5~meV to 3.4~meV depending on the parameter $\eta$.  As expected, our approximation is seen to overestimate the condensation energy and, correspondingly, the interaction energy. However, the difference between the exact and approximate values appears to be significantly smaller than their variations with changing the shape of nanograins. Due to an extremely high sensitivity of energies $\varepsilon_i$ to the grain shape, even within a set of grains with a relatively weak dispersion of shape ($\eta\in [0, 0.06]$), one finds a vast variety of different single-electron energy spectra: from those with a single multiply degenerate energy level in the interaction band (at $\eta=0$) to the spectra with quasi-uniform distributions of levels over the interaction band. As implied by Fig.~\ref{fig1}, the developed approximation provides an adequate description of the interaction energy in all these cases.    

While the results shown in Fig.~\ref{fig1} relate to ultra-small grains, our next goal is to estimate the accuracy of Eq.~(\ref{Eiafin}) in the opposite limit of $n \to \infty$. We perform such an estimate for the ground state of $n$ electron pairs in the half-filled ($u=2n$) interaction band formed by the equidistant single-electron energy levels ($\varepsilon_{i+1}-\varepsilon_{i}=d$ with $d=\hbar\omega_{\rm D}/n$, $i\in {\cal I}$). The exact result for the ground-state interaction energy of such a system at $n \to \infty$ is~\cite{rsd02}
\begin{eqnarray}
E^{\mathrm I}_{\rm g.s.}&&=\hbar\omega_{\rm D} n\left(\sqrt{1+\left[\sinh \left(\frac{1}{\lambda}\right)\right]^{-2}}-1\right) \nonumber\\
&&\approx\hbar\omega_{\rm D} n\times\left\{\begin{array}{l}
\lambda,  \quad \lambda \gg 1 , \\ 
2\exp\left(-{2}/{\lambda}\right), \quad \lambda \ll 1.
\end{array}\right.   
\label{Nlim1}\end{eqnarray}
In the case of $\lambda\to \infty$, our equation~(\ref{Eiafin}) becomes exact resulting in the same value of the ground-state interaction energy as that given by the corresponding expression  of  Eq.~(\ref{Nlim1}). In the case of $\lambda\ll 1$, under the reasonable assumption $c\ll \hbar\omega_{\rm D}$ we obtain from Eq.~(\ref{Eiafin}) the expression   
$ E^{\mathrm I}_{\rm g.s.} \approx c^2n\lambda (2\hbar\omega_{\rm D})^{-1}[1-\lambda\ln (\hbar\omega_{\rm D}/c)]^{-1}$,
which must be minimized with respect to $c$. The minimizing value of $c$ appears to be $c=\hbar\omega_{\rm D} \exp (-1/\lambda+1/2)$, which is indeed much smaller than $\hbar\omega_{\rm D}$. Consequently, the interaction energy takes the value  
\begin{eqnarray}
E^{\mathrm I}_{\rm g.s.} \approx \hbar\omega_{\rm D} n \exp\left(-\frac{2}{\lambda}+1\right),
\label{Nlim3}\end{eqnarray}
which differs from the corresponding result of Eq.~(\ref{Nlim1}) by the factor ${\rm e}/2\approx 1.36$. 

The accuracy of the developed approximation is amazing when taking into account the simplicity of this approximation and the possibility to use it over a wide range of $\lambda$ and $n$ at arbitrary distributions of single-electron energy levels in the interaction band. The accuracy of the results for the magnetic susceptibility, obtained on the basis of our approximation, is analyzed in Appendix B.

\section{Magnetic susceptibility for assemblies of grains}

\subsection{Pairing and parity effects in the magnetic susceptibility}

We start with the magnetic response of parallelepiped-shaped grains similar to those considered in Subsection II.C.3. Within an assembly of grains, grain sizes $l_i$ ($i=x,y,z$) are assumed to vary as $l_i=\langle l_i \rangle(1+r_i\delta)$, where $r_i$ are random numbers ($-1< r_i <1$) and $\delta$ is the amplitude of the relative size deviations. In the further calculations, we take $\langle l_x \rangle : \langle l_y \rangle : \langle l_z \rangle=1:7:7$ and $\delta= 0.025$. The absolute values of $l_i$ ($i=x,y,z$) are determined by a choice of the parameter $K_0$ in the equation $2K_0+P=\langle l_x \rangle \langle l_y \rangle \langle l_z \rangle n_e$, which gives the number of electrons in a grain with volume $\langle l_x \rangle \langle l_y \rangle \langle l_z \rangle$.   

In Fig.~\ref{fig2} we show the normalized magnetic susceptibility $\bar\chi(H,T)$ calculated for two assemblies of even grains. Figures~\ref{fig2}(a) and \ref{fig2}(b) display the results for normal grains ($\lambda=0$) and for superconductor grains ($\lambda=0.22$), respectively. In both cases, the parameter $K_0$ is 2000 and the corresponding average nearest-neighbor level spacing in the interaction band is $\langle d \rangle=3.1$~meV.  

As shown above, at $T=0$, non-zero values of the magnetic susceptibility of a single normal grain correspond to jumps of magnetization, which occur at the magnetic fields $H_k$ given by Eq.~(\ref{field}). For an assembly of grains at $T=0$, the magnetic susceptibility, defined by Eq.~(\ref{sus}), is determined by contributions of grains with $H_k\in (H, H+\Delta H)$, where $k$ takes any allowed value ($0\leq k\leq K-1$). When taking the limit of an infinite ensemble of grains and $\Delta H\to 0$, the magnetic susceptibility $\langle \chi(H,0)\rangle$ is proportional to the probability density of finding $H_k=H$ (with any allowed value of $k$) within this ensemble.  As we mentioned in Ref.~\onlinecite{2}, at $H=0$ parallelepiped-shaped grains are characterized by the Poisson distribution of the spacing between single-electron energy levels. This means that the probability density $w_{m}(s)$ for a particular value $s$ of the spacing between the levels $\varepsilon_{j+m+1}$ ($m=0,1,2,\ldots...$) and $\varepsilon_{j}$ in a grain with the mean nearest-neighbor level spacing $d$ takes the value 
\begin{eqnarray} 
w_{m}^{\rm Poisson}(s)=\frac{1}{m!}\left( \frac{s}{d}\right)^m {\mathrm{exp}}\left(-\frac{s}{d}\right). 
\label{p}
\end{eqnarray} 
Taking into consideration Eq.~(\ref{field}) and assuming $K\gg 1$, the magnetic susceptibility $\chi$ for an (infinite) ensemble of normal grains can be approximated by the expression      
\begin{eqnarray} 
\langle \chi\rangle=\frac{4\mu_{\mathrm B}^2}{d}\sum\limits_{k=0}^{\infty}w_{2k+P}(2\mu_{\mathrm B}H), 
\label{ptot}
\end{eqnarray} 
where the sum over $m$ determines the total probability of finding a spacing $2\mu_{\mathrm B}H$ between two single-particle energy levels, separated from each other by any even (for $P=0$) or any odd (for $P=1$) number of other levels (cp. Refs.~\onlinecite{5a,5}). For an ensemble of even grains with the Poisson distribution of the energy-level spacing, one obtains from Eqs.~(\ref{p}) and (\ref{ptot}):    
\begin{eqnarray} 
\langle \chi\rangle^{\rm Poisson}_{P=0}=\frac{2\mu_{\mathrm B}^2}{d}\left[1+ {\mathrm{exp}}\left(- \frac{4\mu_{\mathrm B}H}{d}\right)\right]. 
\label{susneven}
\end{eqnarray} 
As seen from Fig.~\ref{fig2}(a), the results of our numerical calculations for normal grains at $T\to 0$ basically are in agreement with Eq.~(\ref{susneven}). Deviations of the calculated values $\bar \chi (H,0)$ from the behavior prescribed by Eq.~(\ref{susneven}) are caused by the following two reasons: (i) fluctuations, related to seemingly random distributions of energy levels $\{\varepsilon_i\}$ in individual grains~\cite{2}, are not completely suppressed even for an assembly of as many as 60000 grains; (ii) for our assemblies of parallelepiped-shaped grains, the value $d$ is somewhat different for different grains. 

From Fig.~\ref{fig2}(a), the magnetic susceptibility $\bar\chi(H,0)$ for even normal grains is seen to have a maximum at $H\to 0$. This maximum is due to the fact that the function $w_0^{\rm Poisson}(2\mu_{\mathrm B}H)$, which describes the probability density for the nearest-neighbor spacing to be $2\mu_{\mathrm B}H$, takes its maximum value at $H\to 0$. As distinct from normal grains, in even superconductor grains a transition from the nonmagnetic state with $k=0$ (when all the electrons are on doubly occupied energy levels $\varepsilon_i$) to a magnetic state with $k\geq 1$ always requires a non-zero energy for pair breaking. Therefore, at $T\to 0$ the low-field magnetic susceptibility for even superconductor grains is zero. A non-zero magnetic response appears only at relatively high magnetic fields, when the magnetic energy $2\mu_{\mathrm B}H$ becomes close to the spectroscopic gap in grains. 

In Fig.~\ref{fig2}(b), one finds a well-pronounced maximum of $\bar\chi$ as a function of $H$ at low temperatures. 
In what follows we explain the nature of this maximum. 

First, let us consider grains, where increasing $H$ leads to consecutive increases of $k$ in the ground state by one. 
From Eq.~(\ref{field}), the average value of the magnetic field, which corresponds to the jump  $k\to k+1$, follows to be
\begin{eqnarray} 
\langle H_{k} \rangle =\frac{1}{2\mu_{\mathrm B}} \left[\left(2k+P+1\right)\langle d\rangle - \left< {\Delta_k E_{0k}^{\mathrm{I}}} \right>\right].    
\label{avh}
\end{eqnarray} 
Due to the paring interaction, the values $\langle H_{k} \rangle$ shift by $\left< -{\Delta_k E_{0k}^{\mathrm{I}}} \right>$ towards higher magnetic fields as compared to the case of normal grains. Since the pair breaking energy $-{\Delta_k E_{0k}^{\mathrm{I}}}$ decreases with increasing $k$, the aforementioned shift decreases, too. While in normal grains 
the difference $\langle H_{k+1}-H_{k} \rangle$ between adjacent values $\langle H_{k} \rangle$ takes a constant value $\langle d\rangle/\mu_{\mathrm B}$ for all $k$, in superconductor grains this difference increases with increasing $k$ from $\left(2\langle d\rangle - \left.\langle{\Delta_k^2 E_{0k}^{\mathrm{I}}} \rangle\right|_{k=0}\right)/(2\mu_{\mathrm B})$ for $k=0$ to $\langle d\rangle/\mu_{\mathrm B}$ for $k>n^{(00)}$. 
This means that for $H>\langle H_{0} \rangle$ the density of the values $\langle H_{k} \rangle$ on the magnetic-field axis decreases with rising $H$. Consequently, for $H>\langle H_{0} \rangle$ the magnetic susceptibility of an assembly of superconductor grains should decrease with $H$, approaching the value $\bar\chi=1$ when $H\to \infty$. Since at magnetic fields $H$, which are considerably lower than $\langle H_{0}\rangle$, the magnetic susceptibility of superconductor grains vanishes, we conclude that the afore-described behavior of $\langle H_{k} \rangle$ should result in the appearance of a maximum of $\bar\chi(H,0)$ at $H\sim \langle H_{0}\rangle$.  

Second, it is worth noticing that the expression $\left.{-\Delta_k E_{0k}^{\mathrm{I}}} \right|_{k=0}+\varepsilon_{K+1}-\varepsilon_{K}$, which determines the value $H_0$ for an even grain, coincides with the definition of the spectroscopic gap (cp. Ref.~\onlinecite{2}). As discussed in Ref.~\onlinecite{2}, the pair-breaking energy $\left.-{\Delta_k E_{0k}^{\mathrm{I}}} \right|_{k=0}$ is a decreasing function of the spacing $\varepsilon_{K+1}-\varepsilon_{K}$. Therefore, in superconductor grains, variations of the magnetic field $H_{0}$ with varying the spacing $\varepsilon_{K+1}-\varepsilon_{K}$ are smaller than those in normal grains. In other words, due to the pairing interaction, the values $H_0$, related to various grains, tend to shrink towards $\langle H_{0} \rangle$ (for $H_k$ with $1\leq k<n^{(00)}$, such a trend is much less pronounced, because with increasing $k$ the pair-breaking energy decreases, while variations of the spacing $\varepsilon_{K+k+P+1}-\varepsilon_{K-k}$ increase). The described effect constitutes the second reason for the formation of the maximum of $\bar\chi(H,0)$ at $H\sim \langle H_{0}\rangle$. 

The third reason is related to the grains, where at a certain magnetic field $H_{k,\Delta k}$ the number of doubly occupied levels $\varepsilon_i$ in the ground state abruptly reduces by a value $\Delta k$ larger than one. At $T=0$, each transition of this kind results in an abrupt jump of the magnetization by a relatively large value $2\mu_{\mathrm B}\Delta k$. In view of Eq.~(\ref{hh}), it is obvious that those transitions lead to an increase $\bar\chi(H,0)$ at $H\leq H_k$ [and, correspondingly, in a decrease of $\bar\chi(H,0)$ at $H> H_k$] as compared to the case when $k$ changes by one. Since the very possibility of these transitions is determined by the pairing interaction, their probability generally takes maximum values at $k=0$, when the pairing interaction is the most efficient. Therefore, the contributions of those transitions to $\bar\chi(H,0)$ take maximum values for $H\lesssim H_{0}$ enhancing the magnetic susceptibility at $H\sim \langle H_{0}\rangle$.  
 
With increasing $k$, the number of blocked energy levels $\varepsilon_i$ in the interaction band increases, resulting in a weakening of pairing correlations.  At relatively high magnetic fields, when the number $k$ becomes larger than $n^{(00)}$, the difference between magnetic susceptibilities of superconductor and normal grains disappears. An increase of temperature also tends to reduce the above difference. Nevertheless, even for $k_{\mathrm B}T \sim \langle d\rangle$, this difference is still observable: while for normal grains the magnetic susceptibility $\bar\chi$ monotonously decreases with increasing $H$, for superconductor grains it first increases and then either saturates or reaches a maximum and further decreases towards the saturation value $\bar\chi=1$.  

In Figs.~\ref{fig3}(a) and \ref{fig3}(b), the normalized magnetic susceptibility $\bar\chi(H,T)$ is shown for, respectively, normal and superconductor grains with $P=1$. Regardless of the value of $\lambda$, the magnetization of an odd grain contains a contribution due to the electron, which is not paired in the ground state at $H = 0$. In the case of $T = 0$, the spin of this electron is fully polarized at arbitrarily weak magnetic fields, resulting in a $\delta$-like peak of the magnetic susceptibility at $H = 0$. Apart from this peak, the behavior of the magnetic susceptibility for an ensemble of odd normal grains at $T=0$ is described by the equation 
\begin{eqnarray} 
\langle \chi\rangle^{\rm Poisson}_{P=1}=\frac{2\mu_{\mathrm B}^2}{d}\left[1- {\mathrm{exp}}\left(- \frac{4\mu_{\mathrm B}H}{d}\right)\right], 
\label{susnodd}
\end{eqnarray} 
which can be easily obtained using Eqs.~(\ref{p}) and (\ref{ptot}) with $P=1$. 
At low magnetic fields, the behavior of $\langle \chi\rangle$ is mainly determined by the function $w_1(2\mu_{\mathrm B}H)$ in Eq.~(\ref{ptot}). As follows from Eq.~(\ref{p}), the function $w_1^{\rm Poisson}(2\mu_{\mathrm B}H)$ goes to zero when decreasing $H$. 

For superconductor grains, the behavior of $\bar\chi(H,T)$ is qualitatively similar in the cases of even an odd grains, except for the aforementioned peak of $\bar\chi(H,T)$ related to the presence of an unpaired electron in the ground state of an odd grain at $H = 0$. Due to the presence of this unpaired electron, one of the energy levels $\varepsilon_i$ in the interaction band of an odd grain is always blocked. Therefore, in odd grains, the interaction energies are, on average, lower and effects, related to pairing correlations, are less pronounced as compared to those in even grains. As a result, the maximum of $\bar\chi(H,0)$ at non-zero magnetic fields is somewhat lower for odd grains than for even grains. One can also see that, as compared to the case of even grains, for odd grains this maximum is shifted towards higher magnetic fields. The reason for this shift can be easily understood by inspecting the expression on r.h.s. of Eq.~(\ref{field}) in the cases of $P=0$ and $P=1$.

\subsection{Effect of the bare-level statistics on the magnetic susceptibility}

Since an experimental realization of grain assemblies, which would contain only even (odd) grains, looks problematic, the case of grain assemblies with equally probable values $P=0$ and $P=1$ seems to be important. Figure~\ref{fig4} displays the plots of $\bar\chi(H,T)$ for assemblies, where even and odd parallelepiped-shaped grains are present in equal proportion.  For normal grains, the magnetic susceptibility is seen to be a monotonously decreasing function of $H$. For superconductor grains, the magnetic susceptibility as a function of $H$ has two well-distinguishable maxima, separated by a deep minimum, at low temperatures ($k_{\mathrm B}T \ll \langle d\rangle$) and monotonously increases with magnetic field at $k_{\mathrm B}T \sim \langle d\rangle$. 

In order to examine how sensitive are the obtained results to the single-electron energy-level statistics, let us consider now the case when energy levels $\varepsilon_i$ in a grain follow the Gaussian orthogonal ensemble (GOE) distribution~\cite{rm} with the mean nearest-neighbor level spacing $d$. In order to make the results commensurable with those obtained above, we assume that the value $d$ varies within the assembly of grains in the same way as in the case of parallelepiped-shaped grains. The calculated plots of $\bar\chi(H,T)$ are shown for normal and superconductor grains in Figs.~\ref{fig5}(a) and \ref{fig5}(b), respectively. For a GOE ensemble, probabilities of different energy spectra $\{\varepsilon_i\}$ may differ from each other by many orders of magnitude. Therefore, suppression of fluctuations, related to random distributions of energy levels $\{\varepsilon_i\}$ in individual grains, is not very efficient even at a relatively large number of grains in an assembly. Apart from those relatively large fluctuations, the behavior of $\bar\chi(H,T)$ for superconductor grains is qualitatively the same in the cases of the Poissonian and GOE statistics [cp. Fig.~\ref{fig4}(b) and Fig.~\ref{fig5}(b)]. At the same time, the maximum of $\bar\chi(H,0)$ at non-zero $H$ is lower in Fig.~\ref{fig5}(b) than in Fig.~\ref{fig4}(b). This is because small values of the nearest-neighbor spacing $\varepsilon_{j+1}-\varepsilon_{j}$ are unfeasible in GOE (cf. ``pseudogap''), as distinct from the case of the Poisson distribution of this spacing. Therefore, the interaction energy, which has been demonstrated to play a determinative role in the formation of the maximum of $\bar\chi(H,0)$, is, on average, smaller in the case of the GOE statistics as compared to the case of the Poissonian statistics \cite{2}.   
Nevertheless, at $T\to 0$ the maximum of $\bar\chi(H,0)$ at non-zero $H$ exceeds 1.5, while the calculations, performed in Ref.~\onlinecite{4} for $T=0$ and $\lambda=0.28$, give the height of the maximum as small as 1.02. We relate this discrepancy to the use of purely perturbative results for the interaction energy in Ref.~\onlinecite{4}. It is noteworthy that exact calculations, carried out for a single nanograin with equally spaced energy levels $\varepsilon_{j}$ ($d=6.6$~meV) at temperatures as high as $T \sim 0.5 d/k_{\mathrm B}$, show a maximum of $\chi(H,0)$, which significantly exceeds $1.02 d/\left( 2\mu_{\mathrm B}^2 \right)$ (see Appendix B).    

As mentioned above, for energy levels described by the GOE statistics, small values of interlevel spacing have low probabilities. In this sense, the behaviour of the function $w_0^{\rm GOE}(s)$, which describes the probability density for the nearest-neighbor spacing $s$ in a GOE ensemble~\cite{rm}, is  similar to that of the function $w_1^{\rm Poisson}(s)$. That is why the plots of $\bar\chi(H,T)$, shown in Fig.~\ref{fig5} and calculated for an assembly of both odd and even grains, appear resembling those for odd grains with the Poissonian statistics of single-electron energy levels (cp. Fig.~\ref{fig5} and Fig.~\ref{fig3}). Due to low probabilities of small interlevel spacings in normal grains with the GOE statistics of single-electron levels, a pronounced minimum appears in the low-temperature magnetic susceptibility $\bar\chi$ as a function of magnetic field at $H$ close to zero. In view of this, it is rather the presence of a minimum in the temperature dependence of the zero-field magnetic susceptibility~\cite{dil2000,fal2002} and of the maximum in $\bar\chi(H)$ at $H\sim \langle H_{0}\rangle$ than the presence of a minimum in the low-temperature magnetic susceptibility $\bar\chi(H)$ at low $H$ what can serve to distinguish the magnetic response of superconductor nanograins from that of normal nanograins. 

\subsection{Effect of the grain size and the interaction strength on the magnetic susceptibility}

Figure~\ref{fig6}, where we go back to the case of parallelepiped-shaped grains, shows the magnetic susceptibility $\bar\chi(H,T)$ for an assembly of superconductor grains (both even and odd) with $K_0 =1000$. When comparing Fig.~\ref{fig6} with Fig.~\ref{fig4}(b), which is for grains with two times larger (on average) volume, one can see that with reducing the grain size the peak of $\bar\chi(H,0)$ at nonzero $H$ becomes lower and shifts towards smaller values of $\mu_{\mathrm B}H/\langle d\rangle$. Of course, the absolute value of $H$, which corresponds to this peak, increases with decreasing the grain size (or, equivalently, with increasing $\langle d\rangle$). The effect of decreasing the grain size on $\bar\chi(H,T)$ can be qualitatively understood when taking into consideration the following two circumstances. On the one hand, a decrease of the grain size leads to an increase of the spectroscopic gap \cite{2}. On the other hand, although the pair-breaking energy rises with reducing the grain size (or, equivalently, with increasing $\langle d\rangle$), this rise is slower than an increase of $\langle d\rangle$. Therefore, the relative contribution of the pair-breaking energy to the spectroscopic gap decreases when grains become smaller.               

A comparison of Fig.~\ref{fig7}, where we plot $\bar\chi(H,T)$ for $\lambda=0.15$, with
Fig.~\ref{fig4}(b), which corresponds to the case of $\lambda=0.22$, gives an idea of how the magnetic susceptibility $\bar\chi(H,T)$ of superconductor grains changes with weakening the pairing interaction. When considering the magnetic susceptibility as a function of the dimensionless variables $\mu_{\mathrm B}H/\langle d\rangle$ and $k_{\mathrm B}T /\langle d\rangle$, the effect of decreasing $\lambda$ on $\bar\chi$ looks very similar to that of decreasing the grain size [cp. Figs.~\ref{fig4}(b), \ref{fig6}, and \ref{fig7}]. This similarity is natural, because both a decrease of $\lambda$ and a decrease of the grain size reduce the ratio of the pair-breaking energy to the average nearest-neighbor level spacing $\langle d\rangle$.    

It is worth noting that the behavior of $\bar\chi(H,T)$, described throughout  this section, is sufficiently stable with respect to an increase of the grain-size dispersion $\delta$ within an assembly of grains. As mentioned above, due to an extremely high sensitivity of the energies $\varepsilon_i$ to the grain shape and size, even within a set of grains with a relatively weak dispersion of sizes ($\delta= 0.025$), one finds a vast variety of different single-electron energy spectra. Therefore, a moderate increase of $\delta$ almost does not affect the dispersion of pairing characteristics over an assembly of grains. For the magnetic susceptibility of grains with $\delta=0.2$, our calculations give results, which practically coincide with those in the case of $\delta= 0.025$ (just for the reason of such a close coincidence, we do not show the plots for $\delta=0.2$).

\section{Conclusions}

We have shown that a simple approximation developed here for the interaction energy in superconductor grains is applicable over a wide range of the grain sizes and of the dimensionless interaction parameter $\lambda$ values at arbitrary distributions of single-electron energy levels in the interaction band.
This approximation allows us to analyze the magnetic response of relatively large assemblies of nanograins even at relatively high temperatures and magnetic fields when the thermal energy $k_{\mathrm B}T$ and the magnetic energy $\mu_{\mathrm B}H$ are larger than the mean interlevel spacing in grains, $d$, so that the number of many-electron states, involved in calculations, can exceed $10^5$ per grain. 

The overall profiles of the magnetic susceptibility of superconductor nanograins as a function of magnetic field and temperature are demonstrated to be qualitatively different from those for normal grains. At $T\to 0$, the normalized magnetic susceptibility $\bar\chi(H)$ of superconductor grains is characterized by a well pronounced maximum at a certain value of $2\mu_{\mathrm B}H$, which is close to the width of the spectroscopic gap. Such a maximum or its reminiscence in the form of an increase of $\chi$ with $H$ at low magnetic fields persist even at $k_{\mathrm B}T  \sim d$, both for even and odd grains as well as for their mixtures. The analyzed distinctions between the magnetic response of superconductor grains and that of normal grains are shown to be qualitatively the same both for the GOE statistics of single-electron energy levels in grains and the Poissonian one. This ``universality'' seems important when taking into consideration the fact that an appropriate choice of statistics to describe the energy-level distribution in concrete nanostructures may be not self-evident. The aforementioned features are sufficiently stable also with respect to variations of the average value of the grain size and its dispersion over an assembly of nanograins. These features are shown to be quite pronounced for the ultra-small superconductor grains, where both the spacing $d$ and its variations within an assembly of grains significantly exceed the superconducting gap $\Delta$, so that pairing correlations can hardly be detected through the tunneling spectra. Our results imply that measuring the magnetic-susceptibility profiles, $\chi(H,T)$, can provide a sensitive probe of the pairing interaction in those grains. 

\begin{acknowledgments}
We acknowledge fruitful discussions with K. Yu. Arutyunov and V. V. Moshchalkov. This work has been supported by the GOA BOF UA 2000, IUAP, FWO-V
projects Nos. G.0306.00 G.0274.01, G.0435.03, the W.O.G. WO.025.99 
(Belgium).
\end{acknowledgments}

\appendix
\section{Example: a grain with degenerate single-electron states in the interaction band at $ H= 0$}

Here we analyze the magnetic response of a grain, where in the absence of a magnetic field all the single-electron states in the interaction band have one and same energy: $\varepsilon_{i}=\varepsilon_{K+P}$ for $i\in {\cal I}$. Assuming that the dimensionless interaction parameter is small, $\lambda\ll 1$, we restrict magnetic fields and temperatures under consideration by the inequalities
$\mu_{\mathrm B}H \ll \hbar \omega_{\rm D}$ and $k_{\mathrm B}T \ll \hbar \omega_{\rm D}$. Under these conditions, excitations, which are related to a change of occupation numbers for the single-electron states outside the interaction band, can be neglected when calculating the magnetization of the grain. Therefore, here we consider only many-electron states, where all the energy levels $\varepsilon_{i}$ below the interaction band are doubly occupied, while all the levels above the interaction band are empty.         

Using Eqs.~(\ref{en}) and (\ref{Eis1}), the energy of $2n^{(00)}+P$ electrons, which reside in the interaction band, can be written as
\begin{eqnarray}
E_{gkl}=&& \left( 2n^{(00)}+P\right)\varepsilon_{K+P} -\frac{2\lambda \hbar \omega_{\rm D}}{I}\left( n^{(00)}-k-g\right)
\left( I-P-n^{(00)}-k-g+1\right) \nonumber\\
&&-(2k+P-2l)\mu_{\mathrm B}H. 
\label{AElev}\end{eqnarray}
The degree of degeneracy of the energy level $E_{gkl}$ is described by the expression 
\begin{eqnarray}
J_{gkl}= C^{I}_{2k+P} C^{2k+P}_{l}
\times\left\{\begin{array}{l}
1,  \quad g = 0 , \\ 
\left( C^{I-P-2k}_{g} -C^{I-P-2k}_{g-1}  \right), \quad g \ge 1 \,,
\end{array}\right.  
\label{Adegen}\end{eqnarray}
where the last factor gives~\cite{rom} the number of $(n^{(00)}-k)$-pair states with a given value of $g$. 
For a state with given $k$ and $l$, the spin magnetization is described by Eq.~(\ref{mag0}). Inserting Eqs.~(\ref{mag0}), (\ref{AElev}) and (\ref{Adegen}) into Eq.~(\ref{mag}) and performing summation over $l$, we obtain for the magnetization of the grain:
\begin{eqnarray}
M= \mu_{\mathrm B} \tanh \left( \frac{h}{t} \right) 
\left(\sum_{k=0}^{k_{\rm max}}D_k\right)^{-1} \sum_{k=0}^{k_{\rm max}}(2k+P)D_k \,,  
\label{Mdegen}\end{eqnarray}
where
\begin{eqnarray}
D_k =&&  C^{I}_{2k+P} \left[ 2\cosh \left( \frac{h}{t} \right) \right]^{2k} \exp\left[-\frac{2k(I-P-k+1)}{It} \right] \nonumber\\
&&\times \left\{1+ \sum_{g=1}^{k_{\rm max}-k} \left( C^{I-P-2k}_{g} -C^{I-P-2k}_{g-1}  \right)\exp\left[-\frac{2 g(I-P-2k-g+1)}{It} \right] \right\}\, , 
\label{Dk}\end{eqnarray}
\begin{eqnarray}
k_{\rm max} = \frac12 \left(I-P-\left|I-P-2n^{(00)} \right|\right)\,,
\label{kmax}\end{eqnarray}
$h = \mu_{\mathrm B}H/(\lambda \hbar \omega_{\rm D})$, and $t = k_{\mathrm B}T/(\lambda \hbar \omega_{\rm D})$.

As discussed in Subsection II.B, the magnetization at $T=0$ is determined by the polarization of electron spins in the ground state, which must be found by minimizing the energy $E_{0k0}$ as a function of $k$. From Eq.~(\ref{AElev}) one finds $\Delta_k^2 E_{0k0} = - 4 \lambda \hbar \omega_{\rm D}/I< 0$. This means that the ground-state energy is given by an edge minimum of $E_{0k0}$ at $k=0$ or at $k=k_{\rm max}$. 
From the equation $E_{000}=E_{0k_{\rm max}0}$ we find that the transition from the ``maximal-pairing'' ground state (with $k=0$, $n^{(00)}$ pairs, and magnetization $M_{00}=\mu_{\mathrm B}P$) to the ``maximal-magnetization'' ground state [with $k=k_{\rm max}$, $(n^{(00)}-k_{\rm max})$ pairs, and the magnetization $M_{\rm max}\equiv M_{k_{\rm max}0}=\mu_{\mathrm B}(2k_{\rm max}+ P)$] occurs at the magnetic field 
\begin{eqnarray} 
H_{0,k_{\rm max}} &&=\frac{\lambda \hbar \omega_{\rm D}\left(I-P-k_{\rm max}+1\right)}{\mu_{\mathrm B}I}
\nonumber\\
 &&=\frac{\lambda \hbar \omega_{\rm D}\left(I-P+\left|I-P-2n^{(00)} \right|+2\right)}{2\mu_{\mathrm B}I}.
\label{h0kmax}
\end{eqnarray} 
The field $H_{0,k_{\rm max}}$ is determined by the pair-breaking energy $-\Delta_k E_{0k}^{\mathrm I}$ averaged over $k_{\rm max}$ pairs broken in the course of the above transition. 
Since for the system under consideration the second forward difference $\Delta_k^2 E_{0k0}/ $ coincides with $-\Delta_k^2 E_{0k}^{\mathrm I}$, the inequality $\Delta_k^2 E_{0k0}< 0$, derived above, implies that the energy $-\Delta_k E_{0k}^{\mathrm I}$ decreases with increasing $k$. Therefore, the value $H_{0,k_{\rm max}}$ decreases with increasing $k_{\rm max}$ or, equivalently, with reducing deviations from half filling, which are described by the quantity $\left|I-P-2n^{(00)} \right|$.

The field $H_{0,k_{\rm max}}$ can be considered as the zero-temperature critical magnetic field for the phase transition between the superconducting and normal states of the grain. At $T=0$ this transition is characterized by a $\delta$-like peak of the magnetic susceptibility as a function of $H$. At non-zero temperatures, the aforementioned peak is smoothed out. For $T\neq 0$, we define the ``critical magnetic field'' as the value of $H$, which corresponds to the maximum of the magnetic susceptibility at a given $T$. The plot of the ``critical magnetic field'' versus temperature can be considered as a ``phase boundary'' between the superconducting and normal states of the grain. An example of those ``phase boundaries'', calculated with the use of Eq.~(\ref{Mdegen}) for even grains with $I=50$ and different filling, is shown in Fig.~\ref{fig8}. While the critical magnetic field at $T= 0$ increases with increasing deviations from half filling, the ``critical temperature'' at $H \to 0$ is seen to demonstrate the opposite trend. This is because the ``critical temperature'' at $H \to 0$ is related to the total energy of breaking $k_{\rm max}$ electron pairs rather than to the average breaking energy per pair, which determines $H_{0,k_{\rm max}}$.   

For odd grains with half filling of the interaction band, an analysis of the magnetization at magnetic fields slightly above $H_{0,k_{\rm max}}$ and low temperatures ($k_{\mathrm B}T \ll \lambda\hbar \omega_{\rm D}$) reveals a well-pronounced re-entrant behavior [see Fig.~\ref{fig9}(a)]. With increasing temperature, magnetization, which equals $M_{\rm max}$ at $T=0$, rapidly reduces downwards the value $M_{00}$, characteristic for the ``maximal pairing'' state. This re-entrant behavior is due to the fact that at $k_{\mathrm B}T \ll \lambda\hbar \omega_{\rm D}$ the magnetization in the vicinity of $H_{0,k_{\rm max}}$ is mainly determined by the interplay between the occupation probabilities for the ``maximal magnetization'' and ``maximal pairing'' states. As seen from Eq.~(\ref{Adegen}) in the case of $P=1$ and $2n^{(00)}=I-P$, the energy level $E_{0k_{\rm max}0}$ is non-degenerate, while the level $E_{000}$ is multiply degenerate ($J_{000}=I$).  
Therefore, for $H \approx H_{0,k_{\rm max}}$ and $k_{\mathrm B}T \sim |E_{0k_{\rm max}0}-E_{000}|$, the thermodynamic average magnetization is close to $M_{00}$ even at magnetic fields above $H_{0,k_{\rm max}}$. As seen from Fig.~\ref{fig9}(b), such a re-entrance does not occur in even grains with half filling, where both energy levels $E_{0k_{\rm max}0}$ and $E_{000}$ are non-degenerate. When deviating from half filling, the ratio $J_{0k_{\rm max}0}/J_{000}$ becomes significantly larger than 1. 
As illustrated by Fig.~\ref{fig9}(c), in those cases a distinction of the temperature from $T=0$ results in a rapid increase of magnetization at $H\lesssim H_{0,k_{\rm max}}$ from $M_{00}$ towards $M_{\rm max}$.  
   
When increasing $I$, the ranges of $H$ and $T$, which correspond to the afore-described re-entrant behavior in odd grains with half filling, gradually reduce (see Fig.~\ref{fig_dop}). The reason for such a reduction is as follows. As we have seen above, the re-entrant behavior is related to a dominant role of the energy levels  $E_{000}$ and $E_{0k_{\rm max}0}$ in determining the thermodynamic average magnetization $M$. However, an increase of $I$ results in a rapid increase of degeneracy $J_{gkl}$ for higher excited states and hence in an increasing contribution of those states to $M$ at $T\neq 0$. Consequently, the relative contribution of  the ``maximal pairing'' states to $M$ decreases.
The larger is $I$, the narrower are the range of magnetic fields above $H_{0,k_{\rm max}}$ and the temperature range, where the ``maximal pairing'' states play a determinative role for the thermodynamic average magnetization.     

\section{On the accuracy of the proposed approach}

In order to assess the accuracy of our approximate approach, let us compare the results, obtained within this approach, with those of numerically exact calculations. The latter calculations are time-consuming and can be actually performed only for a relatively small number   
of relatively small grains. In Fig.~\ref{fig10} we compare the approximate and numerically exact results for the zero-temperature magnetic susceptibility of superconductor grains with $ K_0 =1000$. Even for those very small grains the computation time for the exact calculations exceeds that for the corresponding approximate calculations by more than 3 orders of magnitude (the above value dramatically increases with rising grain sizes). Therefore, the number of grains in the exact calculations is limited to the values, which are not yet large enough for a sufficient suppression of fluctuations, related to seemingly random distributions of the energy levels $\{\varepsilon_i\}$ in individual grains. Nevertheless, we can see from Fig.~10 a reasonable agreement between the approximate and exact results. Since our approximate approach somewhat overestimates the interaction energy \cite{2}, the maximum of the approximate curve $\bar\chi(H)$ is higher and lies at stronger $H$ as compared to the exact result. The approximate curve shows also a splitting of this maximum.  At $T\to 0$, a similar splitting is seen also in Figs.~\ref{fig5}(b), \ref{fig6}, and \ref{fig7}. From the analysis of the approximate results for assemblies of grains with a definite parity of the number of electrons, we may draw a supposition that this splitting is related to the parity effect on the position of the maximum of $\bar\chi(H)$ (namely, for odd grains this maximum is shifted toward higher magnetic fields as compared to the case of even grains). However, relatively large fluctuations of the numerically exact curve do not allow us to definitely conclude whether such a splitting should be actually observable or its appearance/magnitude is an artifact of the approximate approach.

Numerically exact calculations of the magnetic susceptibility at non-zero temperatures are practically possible only for single grains. As an example, we consider an even grain with equidistant energy levels $\varepsilon_{j}$ spaced by $d=6.6$~meV. At $k_{\mathrm B}T \sim \langle d\rangle$, calculations of $\chi(H,T)$ for such a grain involve about $5\times 10^5$ many-electron states (this value gives an idea of difficulties in extending the exact calculations to an assembly of grains). In Fig.~\ref{fig11}(a), the results for a normal grain ($\lambda=0$) are shown. Positions of peaks of $\chi(H,0)$ correspond to those given by Eq.~(\ref{field}) for the case of the equidistant energy levels $\varepsilon_{j}$:
\begin{eqnarray} 
H_k=\frac{(2k+1)d}{2\mu_{\mathrm B}}.   
\label{field3}
\end{eqnarray} 
Figures~\ref{fig11}(b) and \ref{fig11}(c) display, respectively, the approximate and exact results for $\lambda=0.22$. Like in the above case of $T=0$, we see a satisfactory agreement between these results. As clear from Figs.~\ref{fig11}(b) and \ref{fig11}(c), in the grain under consideration, the number of doubly occupied energy levels $\varepsilon_{j}$ in the ground state, $K-k$, decreases consecutively by one with increasing $H$. In this grain, the pair-breaking energy $(-{\Delta_k E_{0k}^{\mathrm{I}}})$ and, correspondingly, the difference between the field $H_{k}$ for the superconductor grain and that for the normal grain are significantly larger for $k=0$ than for $k\geq 1$. 
As a result, for the superconductor grain the distance $H_{1}-H_{0}$ between the first two peaks of $\chi(H,0)$ appears significantly smaller than the distances $H_{k+1}-H_{k}$ for the further peaks (with $k\geq 1$). 
This effect is a single-grain analog of the behavior of $\bar\chi(H,0)$ in assemblies of superconductor grains, which is discussed in Subsection III.A: due to the pairing interaction, the magnetic susceptibility of an even superconductor grain almost vanishes at low magnetic fields and takes the maximum values when the magnetic energy $2\mu_{\mathrm B}H$ becomes close to the spectroscopic gap.

\newpage
~

\begin{figure}
\begin{center}
\includegraphics[scale=1.2]{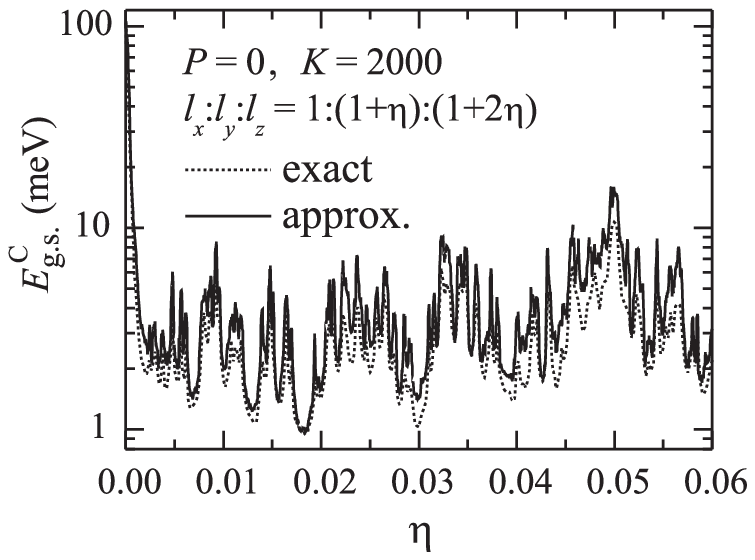}
\end{center}
\caption{Condensation energy, calculated for parallelepiped-shaped grains with the aspect ratio $ l_x: l_y: l_z =1: (1+\eta): (1+2\eta)$, as a function of the parameter $\eta$. The volume of grains is fixed to keep a constant number of electrons in a grain, $2K=4000$. The dimensionless interaction strength, $\lambda$, is taken to be 0.22. The dotted curve represents the numerically exact results~\cite{2}. The solid curve displays the results based on the use of Eq.~(\ref{Eiafin}).}
\label{fig1}
\end{figure}

\begin{figure}
\begin{center}
\includegraphics[scale=0.6]{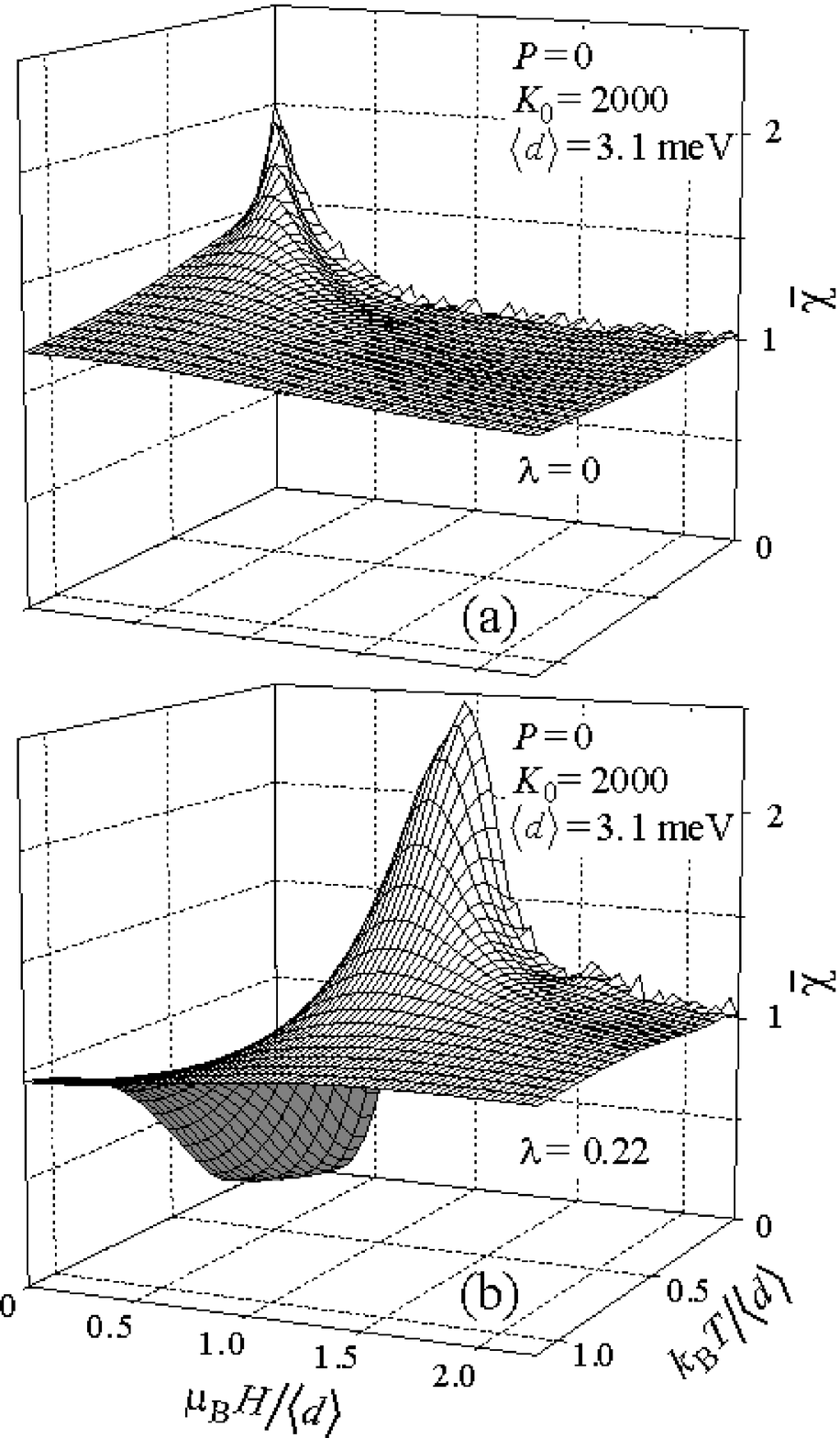}
\end{center}
\caption{Normalized magnetic susceptibility as a function of temperature and magnetic field for two assemblies of parallelepiped-shaped even grains: (a) normal grains ($\lambda=0$) and (b) superconductor grains ($\lambda=0.22$). For each of these assemblies, the total number of grains is $6\times 10^4$, the parameter $K_0$ is 2000, and the average nearest-neighbor level spacing in the interaction band is $\langle d \rangle=3.1$~meV.}
\label{fig2}
\end{figure}

\begin{figure}
\begin{center}
\includegraphics[scale=0.6]{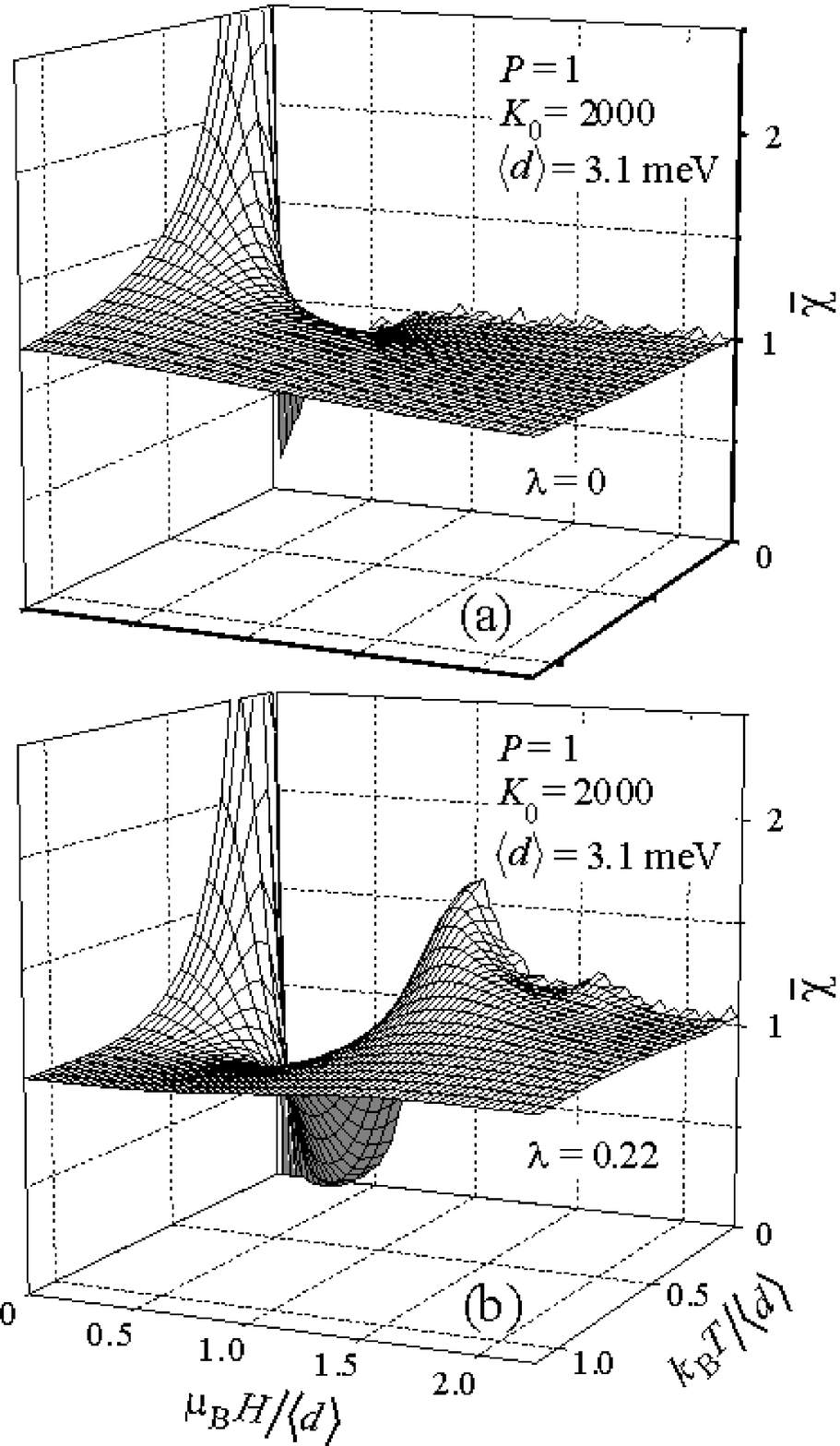}
\end{center}
\caption{Normalized magnetic susceptibility as a function of temperature and magnetic field for two assemblies of parallelepiped-shaped odd grains: (a) normal grains ($\lambda=0$) and (b) superconductor grains ($\lambda=0.22$). For each of these assemblies, the total number of grains is $6\times 10^4$, the parameter $K_0$ is 2000, and the average nearest-neighbor level spacing in the interaction band is $\langle d \rangle=3.1$~meV.}
\label{fig3}
\end{figure}

\begin{figure}
\begin{center}
\includegraphics[scale=0.6]{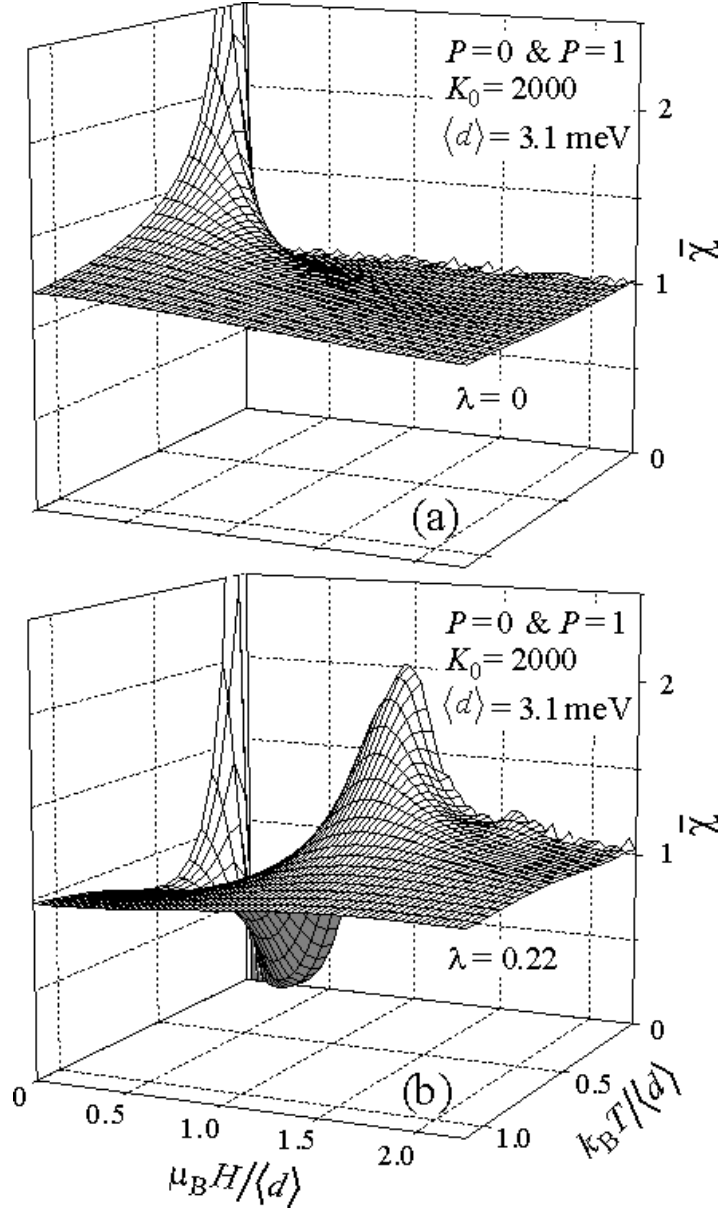}
\end{center}
\caption{Normalized magnetic susceptibility as a function of temperature and magnetic field for two assemblies of parallelepiped-shaped grains: (a) normal grains ($\lambda=0$) and (b) superconductor grains ($\lambda=0.22$). For each of these assemblies, the total number of grains is $1.2\times 10^5$, even and odd grains are mixed in equal proportion, the parameter $K_0$ is 2000, and the average nearest-neighbor level spacing in the interaction band is $\langle d \rangle=3.1$~meV.}
\label{fig4}
\end{figure}

\begin{figure}
\begin{center}
\includegraphics[scale=0.6]{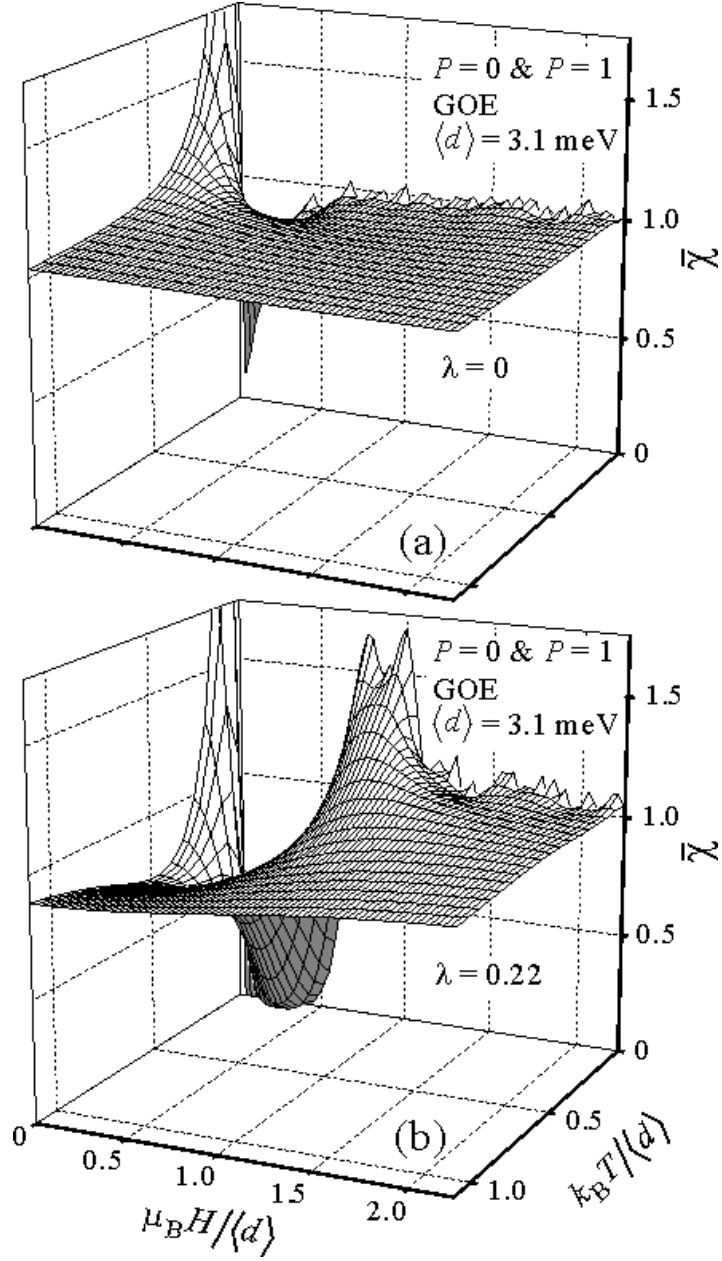}
\end{center}
\caption{Normalized magnetic susceptibility as a function of temperature and magnetic field for two assemblies of grains with the GOE statistics of single-electron energy levels: (a) normal grains ($\lambda=0$) and (b) superconductor grains ($\lambda=0.22$). For each of these assemblies, the total number of grains is $1.4\times 10^6$, even and odd grains are mixed in equal proportion, and the average nearest-neighbor level spacing in the interaction band is $\langle d \rangle=3.1$~meV.}
\label{fig5}
\end{figure}

\begin{figure}
\begin{center}
\includegraphics[scale=0.6]{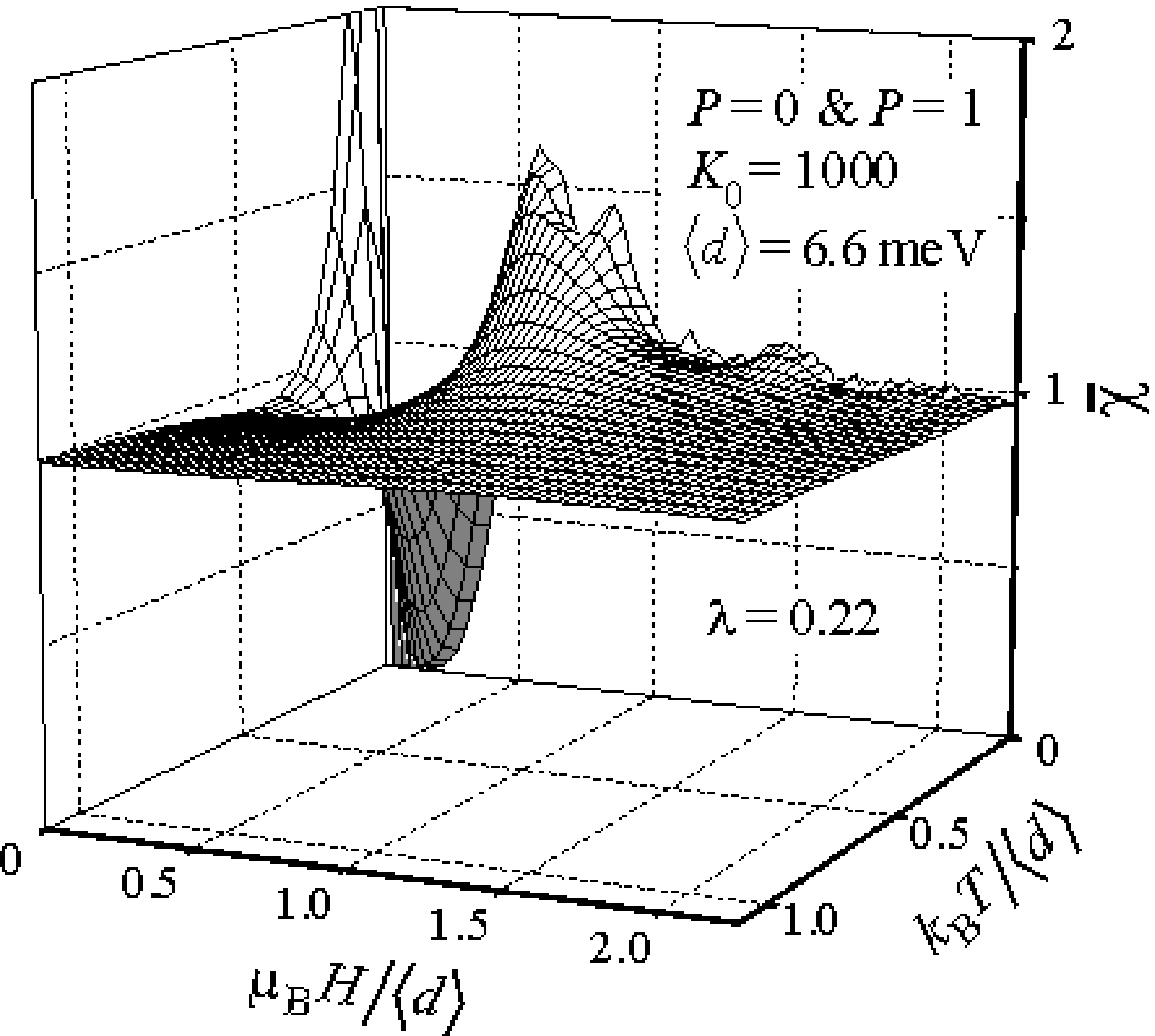}
\end{center}
\caption{Normalized magnetic susceptibility as a function of temperature and magnetic field for an assembly of parallelepiped-shaped superconductor grains with $\lambda=0.22$. The total number of grains is $1.2\times 10^5$, even and odd grains are mixed in equal proportion, the parameter $K_0$ is 1000, and the average nearest-neighbor level spacing in the interaction band is $\langle d \rangle=6.6$~meV.}
\label{fig6}
\end{figure}

\begin{figure}
\begin{center}
\includegraphics[scale=0.6]{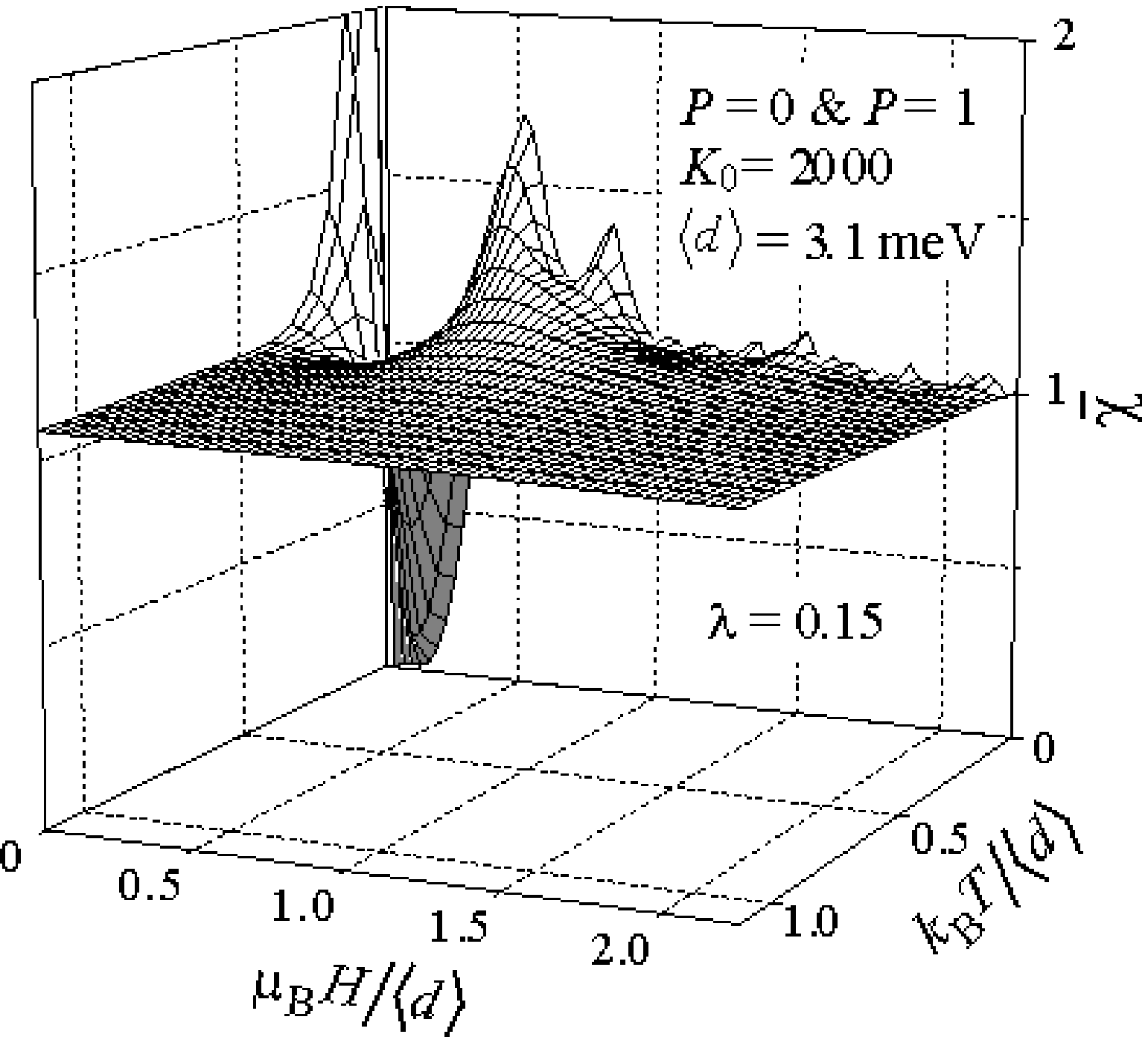}
\end{center}
\caption{Normalized magnetic susceptibility as a function of temperature and magnetic field for an assembly of parallelepiped-shaped superconductor grains with $\lambda=0.15$. The total number of grains is $1.2\times 10^5$, even and odd grains are mixed in equal proportion, the parameter $K_0$ is 2000, and the average nearest-neighbor level spacing in the interaction band is $\langle d \rangle=3.1$~meV.}
\label{fig7}
\end{figure}

\begin{figure}
\begin{center}
\includegraphics[scale=1.2]{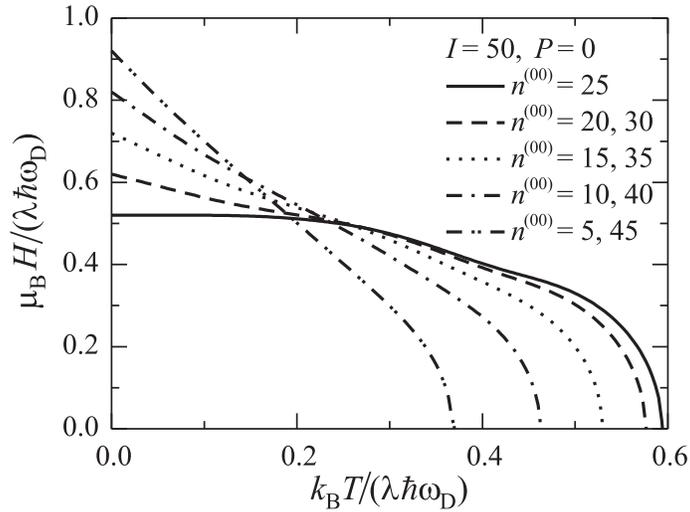}
\end{center}
\caption{``Phase boundaries'' for even superconductor grains with $2n^{(00)}$ electrons distributed over $2I$ single-electron states in the interaction band, all these states possessing the same energy in the absence of a magnetic field.}
\label{fig8}
\end{figure}

\begin{figure}
\begin{center}
\includegraphics*{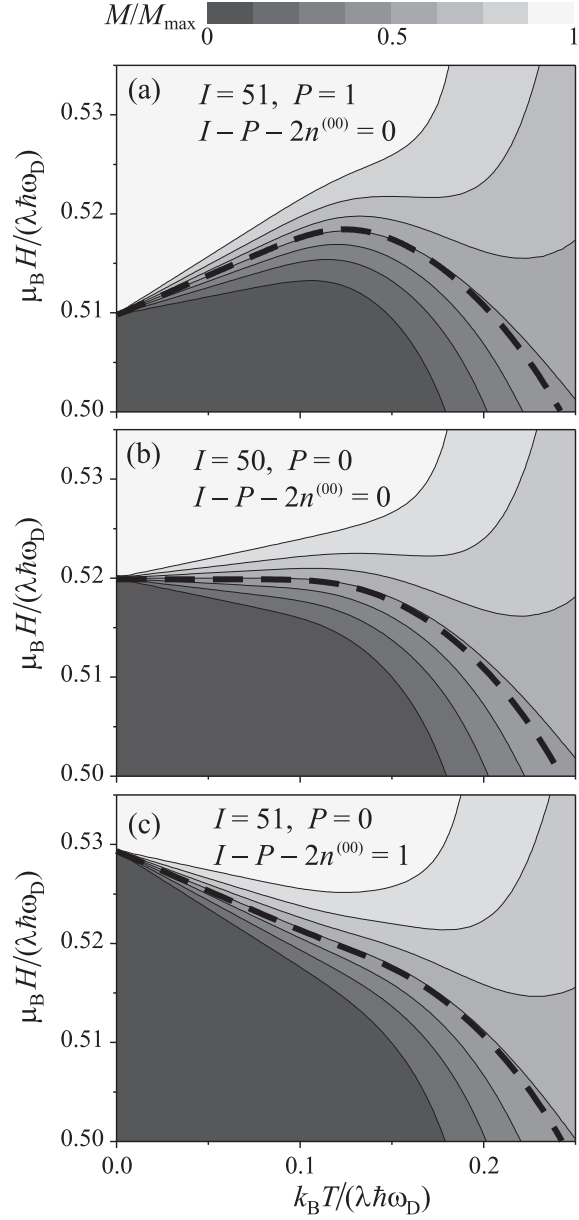}
\end{center}
\caption{Magnetization of superconductor grains, where $2I$ single-electron states in the interaction possess the same energy at $H=0$, as a function of temperature and magnetic field. The number of electrons in the interaction band is $2n^{(00)}+P$. Heavy dashed lines show the corresponding ``phase boundaries''.  Panel (a): an odd grain with half filling of the interaction band. Panel (b): an even grain with half filling of the interaction band. Panel (c): an even grain with a slight deviation from half filling of the interaction band.}
\label{fig9}
\end{figure}

\begin{figure}
\begin{center}
\includegraphics[scale=1.2]{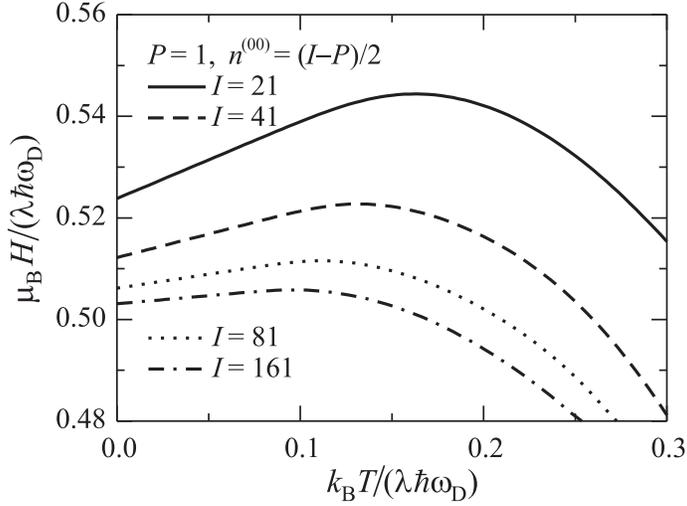}
\end{center}
\caption{``Phase boundaries'' for odd superconductor grains with different number $2I$ of single-electron states in the half-filled interaction band, all these states possessing the same energy in the absence of a magnetic field.}
\label{fig_dop}
\end{figure}

\begin{figure}
\begin{center}
\includegraphics[scale=1.2]{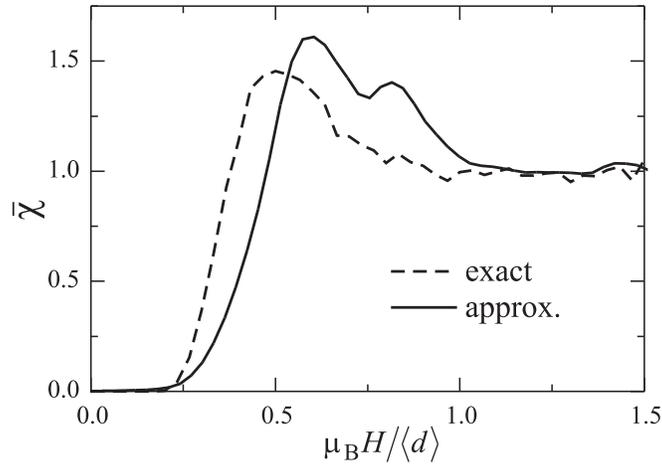}
\end{center}
\caption{Normalized magnetic susceptibility as a function of magnetic field for parallelepiped-shaped superconductor grains with $\lambda=0.22$ at $T=0$.
The dashed curve shows the numerically exact results for an assembly of $4\times 10^4$ grains. The solid curve displays the results, based on the use of Eq.~(\ref{Eiafin}), for an assembly of $6\times 10^6$ grains. In both cases, even and odd grains are mixed in equal proportion, the parameter $K_0$ is 1000, and the average nearest-neighbor level spacing in the interaction band is $\langle d \rangle=6.6$~meV.}
\label{fig10}
\end{figure}

\begin{figure}
\begin{center}
\includegraphics[scale=0.65]{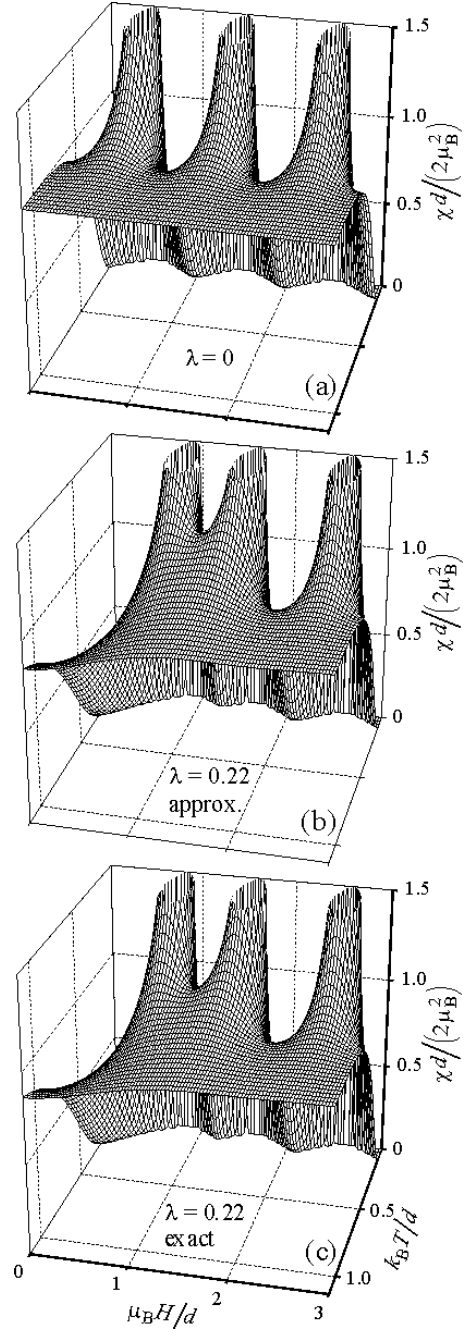}
\end{center}
\caption{Normalized magnetic susceptibility as a function of temperature and magnetic field for single even grains with the equidistant single-electron energy levels spaced by $d=6.6$~meV. Panel (a) shows the results for a normal grain ($\lambda=0$). For a superconductor grain with $\lambda=0.22$, the approximate results, based on the use of Eq.~(\ref{Eiafin}), and the numerically exact results are displayed on panels (b) and (c), respectively.}
\label{fig11}
\end{figure}

\end{document}